\documentclass[useAMS]{mn2e}
\usepackage{amssymb,amsmath,psfig,times}
\usepackage{graphicx}
\def\gsim{ \lower .75ex \hbox{$\sim$} \llap{\raise .27ex \hbox{$>$}} }
\def\lsim{ \lower .75ex\hbox{$\sim$} \llap{\raise .27ex \hbox{$<$}} }

\def\beq{\begin{equation}}
\def\eeq{\end{equation}}

\def\sw{{\it Swift}}
\def\fe{{\it Fermi}}
\def\ba{BATSE}
\def\cgro{{\it CGRO}}

\def\ep{$E_{\rm p}$}

\def\epo{$E^{\rm obs}_{\rm p}$}

\def\liso{$L_{\rm iso}$}
\def\eiso{$E_{\rm iso}$}
\def\ama{$E_{\rm p}-E_{\rm iso}$}
\def\yone{$E_{\rm p}-L_{\rm iso}$}
\def\ghi{$E_{\rm p}-E_{\gamma}$}
\def\epof{$E^{\rm obs}_{\rm peak}-F$}

\def\th{$\theta_{\rm jet}$}

\def\thv{$\theta_{\rm view}$}
\def\tjet{$t_{\rm break}$}
\def\tpeak{$t_{\rm peak}$}
\def\egamma{$E_{\gamma}$}
\def\flim{$P_{\rm lim}$}
\def\G{$\Gamma_{0}$}

\def\lisocom{$L'_{\rm iso}$}
\def\eisocom{$E'_{\rm iso}$}
\def\egcom{$E'_{\gamma}$}

\def\epcom{$E'_{\rm p}$}

\title[Characteristic \th\ and \G\ in GRBs]
{The faster the narrower: characteristic bulk velocities and jet opening angles of Gamma Ray Bursts}
\author[G. Ghirlanda et al.]
{G. Ghirlanda$^{1}$\thanks{E--mail:giancarlo.ghirlanda@brera.inaf.it}, 
G. Ghisellini$^{1}$, R. Salvaterra$^{2}$, L. Nava$^{3}$, D. Burlon$^{4}$, G. Tagliaferri$^{1}$,\newauthor 
S. Campana$^{1}$,  P. D'Avanzo$^{1}$, A. Melandri$^{1}$\\
$^{1}$INAF -- Osservatorio Astronomico di Brera, via E. Bianchi 46, I-23807 Merate, Italy\\
$^{2}$INAF - IASF Milano, via E. Bassini 15, I-20133 Milano, Italy \\
$^{3}$APC Universit\'e Paris Diderot, 10 rue Alice Domon et Leonie Duquet, F-75205 Paris Cedex 13, France\\
$^{4}$Sydney Institute for Astronomy, School of Physics, The University of
Sydney, NSW 2006, Australia\\
}
\begin{document}

\date{}


\maketitle

\label{firstpage}

\begin{abstract}
The jet opening angle \th\ and the bulk Lorentz factor \G\
are crucial parameters for the computation of the  energetics
of Gamma Ray Bursts (GRBs).
From the $\sim$30 GRBs with measured  \th\ or \G\ it is known that: 
(i) the real energetic \egamma, obtained by correcting the isotropic equivalent energy \eiso\ for 
the collimation factor $\sim\theta^2_{\rm jet}$, is clustered around $10^{50}$--$10^{51}$ erg 
and it is correlated with the peak energy \ep\ of the prompt emission and 
(ii) the comoving frame \epcom\ and \egcom\ are clustered around typical values. 
Current estimates of \G\ and \th\ are based on incomplete data samples and their observed distributions
could be subject to biases.
Through a population synthesis code
we investigate whether 
different assumed intrinsic distributions of \G\ and \th\ can reproduce a set of observational constraints
Assuming that all bursts have the same \epcom\ and \egcom\ in the comoving frame, 
we find that \G\ and \th\ cannot be distributed as single power--laws.
The best agreement between our simulation and the available data
is obtained assuming (a) log--normal distributions for \th\ and \G\ and (b) an intrinsic
relation between the peak values of their distributions, i.e \th$^{2.5}$\G=const.
On average, larger values of \G\ (i.e. the ``faster" bursts) correspond 
to smaller values of \th\ (i.e. the ``narrower"). 
We predict that $\sim$6\% of the bursts that point to us should not show any jet break in their 
afterglow light curve since they have $\sin\theta_{\rm jet}<1/$\G. 
Finally, we  estimate that the local rate of 
GRBs is $\sim$0.3\%  of all local SNIb/c and $\sim$4.3\% of local hypernovae,
i.e. SNIb/c with broad--lines.
\end{abstract}

\section{Introduction}

Gamma Ray Bursts (GRBs) have extremely high energetics. The isotropic 
equivalent energy \eiso, released during the prompt phase, is distributed over four orders of 
magnitudes in the range 10$^{50-54}$ erg. \eiso\ correlates with \ep, i.e. the peak of the $\nu F_{\nu}$ spectrum  
(Amati et al. 2002, 2009): $E_{\rm p}\propto E_{\rm iso}^{0.5}$. 
This holds for long duration GRBs. 
A similar correlation exists between 
the isotropic equivalent luminosity \liso\ and \ep\ (Yonetoku et al. 2004) obeyed also by short events (Ghirlanda et al. 2009). 
The scatter of the data points 
around the \ama\ correlation, modeled with a Gaussian, has a dispersion  
$\sigma_{\rm sc}=$ 0.23 dex (see e.g. Nava et al. 2012 for a recent update of these correlations). 
This dispersion is much larger than the average statistical error $\bar{\sigma}_{E_{\rm iso}}=0.06$ dex and 
$\bar{\sigma}_{E_{\rm peak}}=0.10$ dex associated with \eiso\  and \ep, respectively.

Since \eiso\ is computed assuming that GRBs emit isotropically, it is only a proxy of the real GRB energetic. 
GRBs are thought to emit their radiation within a jet of opening angle \th. If the jet opening angle \th\  
is known, the
{\it true energy} \egamma$\simeq$\eiso $\theta_{\rm jet}^2$  
and the {\it true GRB rate} can be estimated (Frail et al. 2001). 

The estimate of \th\  is made possible by the measure of the jet break time 
\tjet, typically observed between 0.1 to $>$10 days in the afterglow optical light curve. 
Although \th\ has been measured only for $\sim$30 GRBs (Ghirlanda et al. 2007) it shows that: 
\begin{enumerate}
\item \egamma\ clusters around $10^{50}$ erg with a small dispersion (Frail et al. 2001; but see 
Racusin et al. 2009; Kocevski \& Butler 2008);
 
\item  \egamma\ is tightly correlated with \ep\ (Ghirlanda, Ghisellini \& Lazzati 2004; Ghirlanda et al. 2007) 
with a scatter $\sigma_{\rm sc}= 0.07$ dex  (consistent with the average statistical error 
$\bar{\sigma}_{E_{\gamma}}=\bar{\sigma}_{E_{\rm p}}\simeq 0.1$ dex associated with 
\egamma\ and \ep);

\item  the true rate of local GRBs ranges from $\sim$ 250 Gpc$^{-3}$ yr$^{-1}$ (e.g. Frail et al. 2001)
to $\sim$ 33 Gpc$^{-3}$ yr$^{-1}$ (Guetta, Piran \& Waxman 2005). 
These different values are mainly due to
the different values assumed for the collimation factor $f\propto \theta_{\rm jet}^{-2}$.
The true GRB rate can be compared with the local rate of SN Ib/c 
(e.g. Soderberg 2006; Guetta \& Della Valle 2007; Grieco et al. 2012), 
i.e. the candidate progenitors of long GRBs, and allows to estimate  
the rate of orphan afterglows (e.g. Guetta et al. 2005). 
\end{enumerate}

The \ama, \yone\ and \ghi\ correlations could enclose some underlying feature of the GRB emission 
mechanism (e.g.  Rees \& Meszaros 2005;  Ryde et al. 2006; Thompson 2006; Giannios \& Spruit 2007; Thompson, 
Meszaros \& Rees 2007; Panaitescu 2009), of the GRB jet structure (e.g. Yamazaki, Ioka \& Nakamura 2004; 
Eichler \& Levinson 2005; Lamb, Donaghy \& Graziani 2005; Levinson \& Eichler 2005) or of the progenitor 
(e.g. Lazzati, Morsony \& Begelman 2011). An intriguing application of these correlations is the use of GRBs as standard candles 
(Ghirlanda, Ghisellini \& Firmani 2005; Firmani et al. 2005; Amati et al. 2009). 

The presence of outliers of the \ama\ correlation in the \cgro/\ba\ GRB population 
(Band \& Preece 2005; Nakar \& Piran 2005; Shahmoradi \& Nemiroff 2011) and in the \fe/GBM burst sample 
(Collazzi et al. 2012) and the presence of possible instrumental biases (Butler et al. 2007; Butler, 
Kocevski \& Bloom 2009; Kocevski 2012) caution about the use of these correlations either for deepening 
into the physics of GRBs and for cosmological purposes. 
Although instrumental selection effects are present, it seems that  they cannot produce the correlations we see 
(Ghirlanda et al. 2008; Nava et al. 2008; Ghirlanda et al. 2012b). Moreover, a correlation between \ep\ and \liso\ 
is present within individual GRBs as a function of time (Firmani et al. 2009;
Ghirlanda et al. 2010; 2011; 2011a),  suggesting that the 
radiative process(es) might be the origin of the \yone\ correlation. Despite these studies, the spectral 
energy correlations of GRBs and their possible applications are still a matter of intense debate. 

A new piece of information recently added to the puzzle is that the GRB energetics 
($E_{\rm iso}^\prime$, $L_{\rm iso}^\prime$\ and $E_{\rm p}^\prime$) appear nearly similar in the comoving frame 
(Ghirlanda et al. 2012 -- G12 hereafter). 
To measure these comoving quantities\footnote{Primed quantities are in the comoving frame of the source.} we have to know the bulk Lorentz factor \G,
that can be estimated  through the measurement of the peak time \tpeak\ of the afterglow light curve.
G12 could estimate \G\ in 30 long GRBs with known $z$ and well defined energetics,
finding that:
\begin{enumerate}
\item \eiso(\liso)$\propto\Gamma_0^2$ and \ep$\propto$\G; 
\item the comoving frame \eisocom$\sim$3.5$\times10^{51}$ erg (dispersion 0.45 dex), 
\lisocom$\sim$5$\times10^{48}$ erg s$^{-1}$ (dispersion 0.23 dex) and \epcom$\sim$6 keV (dispersion 0.27 dex). 
\end{enumerate}
These results imply that the \ama\ and \yone\ correlation are a sequence of different \G\ factors 
(see also Dado, Dar \& De Rujula 2007).

The \th\ values of GRBs are known only for a couple of dozens of bursts (Ghirlanda et al. 2007). 
\th\ appears distributed as a log--normal with a typical \th$\sim3^\circ$ (Ghirlanda et al. 2005). 
By correcting the isotropic comoving frame energy \eisocom\ by this typical jet opening angle, 
the comoving frame true energy \egcom\ results $\sim 5\times 10^{48}$ erg. 
In G12 we also argued that in order to have consistency between the \ghi\ and the \ama\ correlations
one must require $\theta_{\rm jet}^2\Gamma_0$ = constant. 
A possible anti--correlation between $\theta_{\rm jet}$ and $\Gamma_0$ is predicted by 
models of magnetically accelerated jets (Tchekhovskoy, McKinney \& Narayan 2009; Komissarov, Vlahakis \& Koenigl  2010) 
but, at present, only 4 GRBs have an estimate of \th\ and \G\ and well constrained spectral properties.

The measure of \th\ relies on the measure of \tjet, that in turn
requires the follow up of the optical afterglow emission up to a few days after the burst explosion 
(Ghirlanda et al. 2007).
The measurement of \tjet\ is difficult, not only because it requires a large investment of telescope time,
but also because several \tjet\ are chromatic (contrary to what predicted; but see Ghisellini et al. 2009), 
and the jet break can be a smooth transition whose measurement  requires an excellent sampling 
of the afterglow light curve (e.g. Van Eerten et al. 2010, 2011). 
Another complication is that the 
early afterglow emission is characterized by several breaks.
For instance, the end of the plateaux phase typically 
observed in the X--ray light curves, if  misinterpreted as a jet break,  biases the \th\ distribution 
towards small values of \th\ (Nava, Ghisellini \& Ghirlanda 2006). 
Finally, the measure of large \th\ is complicated by the faintness of the afterglow and its possible 
contamination by the host galaxy emission and the supernova associated to the burst. 
Several observational biases could shape the observed \th\ distribution. Among these the fact that 
more luminous bursts (i.e. those more easily detected) should have the smallest jet opening 
angles. For all these reasons the {\it observed} distribution of \th\ might not be representative of the real 
distribution of GRBs jet opening angles. 

The distribution of \G\ is centered around \G=65 (130) in the case of a wind (uniform) density distribution 
of the circum--burst medium. 
The distribution of \G\ is broad and extends between \G$\sim$20 and \G$\sim$800.
These results are still based on a sample of only 30 GRBs (G12). 
The difficulties of early follow--up of the optical afterglow emission could prevent the measure of very 
large \G\ on the one hand, while the possible contamination by flares (Burrows et al. 2005; Falcone et al. 2007) or by other (non afterglow) 
emission components (e.g. Ghisellini et al. 2010) at intermediate times could prevent the estimate of the 
low--end of the \G\ distribution.  
One could argue if GRBs can have \G\ of a few. While there are some hints that GRB060218 
should have \G$\sim$5 (Ghisellini et al. 2006)  the classical compactness argument, for typical GRB parameters (e.g. Piran 1999), 
requires that  \G$\ge$100-200. This argument  was successfully applied to few bursts observed up to GeV energies by LAT on board Fermi 
(e.g. Abdo et al. 2009, Ghirlanda et al. 2009) to derive lower limits of several hundreds on \G. If, instead, 
the highest energy photon detected has an energy of say $E_{\rm max}\sim3$ MeV, the lower limit derived from the classical compactness 
argument would be \G$>$ a few (i.e. $\sim 2E_{\rm max}/m_{e} c^{2}$). 
Therefore, also in the case of \G, the {\it observed} distribution, derived with still few events, could be not representative of the 
real distribution of this parameter.  

The main aim of this paper is to constrain the distribution of \G\ and \th\ in GRBs using the available independent constraints. 
This aim can be translated into a simple question: do \th\ and \G\ follow power law distributions  
or do they follow some kind of peaked distribution (e.g. a broken power law or a log--normal)? 
In both cases the resulting distributions could be different from the observed ones since some selection effect (as discussed above) might prevent to measure very low and/or high values of \G\ and \th.  Another scope of the present paper is to test which is (if any) the relation between \th\ and \G. A relation \th$^2$\G=const was assumed in G12 to explain the spectral energy correlations and a similar relation seems to arise from  numerical simulations of jet accelerations (Tcheckolskoy et al. 2012). Here we use several observational constraints and test whether there is a \th$^a$\G=const relation and try to constrain its exponent $a$.  One important effect that we consider in this paper for the first time is the collimation of the burst radiation when \G\ is small. In general we are led to think that given a value of the collimation corrected energy \egamma, the corresponding isotropic equivalent energy is \eiso$\sim$\egamma/\th$^2$. This is true if the beaming of the radiation is ``dominated'' by the jet opening angle, i.e. 1/\G$\le \sin$\th. However, GRBs with very low \G\ could have 1/\G$\ge \sin$\th\ and in this case the isotropic equivalent energy is determined by \G\ (i.e. \eiso$\sim$\egamma\G$^2$ - see \S. 2.2) rather than by \th. This effect, introduces a limit ($\propto$\eiso$^{1/3}$) in the classical \ama\ plane (Fig.1) accounting for the absence of bursts with intermediate/low \ep\ and large values of \eiso. This limit can also partly account for the problem of ``missing jet breaks" since these bursts with 1/\G$\ge \sin$\th\ should not show any jet break in  their afterglow light curve (\S4.5). 


We rely on a GRB population synthesis code that we have recently adopted to explore the issue of 
instrumental selection biases on the \yone\ correlation (Ghirlanda et al. 2012b). 

The simulation steps are described in \S2 while the observational constraints that we aim to 
reproduce are outlined in \S3. 
In \S4 we present our results. We summarize and discuss our findings in \S5.
Throughout the paper a standard flat universe  with $h=\Omega_{\Lambda}=0.7$ is assumed.

\section{Population synthesis code}

\begin{figure*}
\hskip -1.2truecm
\psfig{file=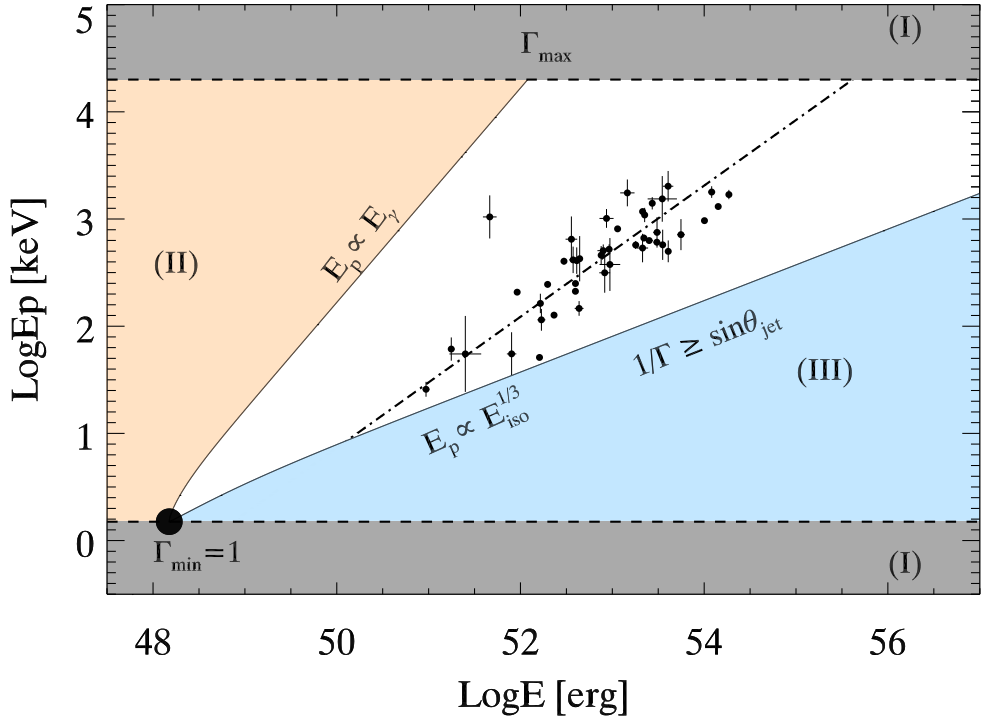,width=16cm}
\vskip -0.3 cm
\caption{
Rest frame plane of GRB energetics. 
The large black dot corresponds to the main assumption of our simulations, i.e. 
that all bursts have similar comoving frame \epcom=1.5 keV and \egcom=1.5$\times 10^{48}$ erg. 
Assigning a certain \G\ to the burst, this moves along the line
$E_{\rm p}\propto E_\gamma$.
Since $\Gamma_0>1$ and we assume a maximum \G\ of 8000, regions (I) are forbidden.
Since all our simulated bursts have \th$\leq$90$^\circ$, they cannot lie in region (II). 
When \G\ is small, the beaming cone $\sim 1/$\G\ can become wider than the
aperture of the jet. 
In this case the isotropic equivalent energy becomes  
\eiso=\egamma$(1+\beta_0)\Gamma_0^2$, that is smaller than the energy calculated through 
\eiso=\egamma/(1--cos\th). 
This introduces a limit $E_{\rm p}\propto E_{\rm iso}^{1/3}$ and bursts cannot lie on the 
right of this limit. Consequently, region (III) is forbidden.
The black dots correpond to the real GRBs of the \sw\ complete sample. 
The fit to the \sw\ complete sample is shown by the dot--dashed line. }
\label{fg0}
\end{figure*}

So far, the approach adopted in studying the spectral--energy correlations and the distributions 
of \th\ or \G\ was 
(i) to derive the collimation corrected \ghi\ correlation
by correcting the isotropic energy \eiso\ 
for the collimation factor $\propto\theta_{\rm jet}^2$ (e.g. Ghirlanda et al. 2004),  or 
(ii) to derive the comoving frame properties of GRBs by correcting, for the \G\ factor, the 
isotropic values \eiso, \liso\ and \ep\ (G12). 

In this paper, we tackle the problem from the opposite side and jointly work with \th\ and \G: 
we assume that GRBs have all the same comoving frame \epcom\ and \egcom\ and  simulate GRB samples with 
different distributions of \G\ and \th. 
This produces a population of GRBs with known energetics 
\eiso, peak energy \ep\ and observer frame fluence $F$ and peak flux $P$. 
We would like to stress that our main assumption (same \epcom\ and \egcom\ for all burst)
is a crude simplification. 
Nevertheless, our assumption can work if the real \epcom\ and \egcom\
distributions  are indeed narrower that the distributions
of the corresponding observed quantities. Recently, Giannios 2012 have shown that in photospheric models a 
comoving frame peak energy \epcom$\sim$1.5 keV is expected.

The observational constraints that we aim to reproduce (see \S3) are: 
(i) the rate of GRBs observed by \sw/BAT, \cgro/\ba\ and \fe/GBM, 
(ii) the \ama\ correlation defined by the complete sample of \sw\ bright bursts 
(Salvaterra et al. 2012; Nava et al. 2012) and 
(iii) the fluence and peak flux distributions of the population of bursts detected by \fe/GBM  
(Goldstein et al. 2012) and \cgro/\ba\ (Meegan et al. 1998).

Note that, since one of the aims of the present paper is to constrain the distributions of \th\ and \G\ we cannot adopt the 
observed ones (discussed in the introduction) as  constraints, otherwise we would fall into a circular argument. The distributions 
of \G\ and \th\ that we assume in our simulations (power law, broken power law, log--normal) have all their characteristic parameters (slope, normalization, break values, width etc.) free to vary. These parameters are what we aim to constrain through our population synthesis code.

In Fig. \ref{fg0} we show the rest frame peak energy \ep\ versus the total energy 
$E$ (where $E$ here is 
generically used to indicate an energy, either isotropic or collimation corrected). 
We highlight different regions 
(I, II and III) that are useful to explain the simulation steps (\S2.1). 
This plane will be one of our observational constraints: 
in Fig. \ref{fg0} we show (black filled points) the \sw\ complete sample of bursts 
(Salvaterra et al. 2012; Nava et al. 2012) 
which we aim to reproduce through our simulations.

\subsection{Simulation steps}
Our starting assumption is that all GRBs have the same 
comoving frame \epcom=1.5 keV and \egcom=1.5$\times10^{48}$ erg. 
This is shown by the black circle in Fig. \ref{fg0}. 
G12 find that 
\lisocom$\sim$const and that the observed duration $T_{90}$ does not depend on \G.
Therefore, in the comoving frame, $T_{90}^\prime \propto \Gamma_0 T_{90} \propto \Gamma_0$.
It follows that \egcom=\lisocom$T'_{90}\theta_{\rm jet}^2$ 
is also constant if, as discussed in G12, $\theta_{\rm jet}^2$\G=const. 
Although some dispersion of the values of \epcom\ is present in the sample of G12, the value 
of \epcom\ that we assume here is consistent at the 2$\sigma$ level of confidence with the 
distribution of values reported in G12 for the wind density ISM. 


The main steps of our simulation are:
\begin{enumerate}

\item we simulate a population of GRBs distributed in redshift $z$ between $z=0$ and $z=10$
according to the GRB formation rate (GRBFR) $\psi(z)$.
This is formed by two parts: $\psi(z)=e(z) R(z)$.
The first term is a cosmic evolution term, while $R(z)$ 
is taken from Li (2008) (which extended to higher 
redshifts the results of Hopkins \& Beacom 2008):
\begin{equation}
R(z)=\frac{0.0157+0.118z}{1+(z/3.23)^{4.66}}
\label{z}
\end{equation}
$R(z)$ is in units of ${\rm M}_{\sun}$ ${\rm yr}^{-1}$  ${\rm Mpc}^{-3}$. 
Concerning $e(z)$, Salvaterra et al. (2012) derived the 
luminosity function of GRBs by jointly fitting the redshift distribution of a complete sample of bright 
GRBs detected by \sw\  and the count distribution of a larger sample of \ba\ bursts. 
They found that either the evolution of the luminosity function or the evolution of the density of GRBs is 
required in order to account for these data sets. 
We assume the same term $e(z)=(1+z)^{1.7}$ found by S12.   

\item We assign to each GRB a bulk Lorentz factor \G\ extracted from a specified distribution, in 
the range [1, 8000]. 
The upper limit ($\Gamma_{\rm 0, max}=8000$) is somewhat arbitrary,
but large enough to encompass all the values of \G\ estimated so far, and in particular 
the large values derived for the few GRBs detected by the LAT instrument on board \fe, 
if the GeV emission is interpreted as afterglow (Ghisellini et al. 2010).

For each simulated burst the rest frame peak energy \ep\ and the energy \egamma\ are (see G12):
\begin{equation}
E_{\rm peak}=E'_{\rm peak}\frac{5\Gamma_{0}}{5-2\beta_{0}} \, \, ; \, \, E_{\gamma}=E'_{\gamma}\Gamma_{0}
\end{equation}
where \G=$1/(1-\beta_{0}^{2})^{1/2}$. 
The simulated bursts define a correlation between \ep\ and \egamma:
\begin{equation}
E_{\rm peak} = \frac{E'_{\rm peak}}{E'_{\gamma}} 
\frac{5 E_{\gamma}}{5-2\beta_{0}} \propto \frac{E_{\gamma}}{5-2\beta_{0}}
\label{lghi}
\end{equation}
for $\beta_{0}\sim1$ this corresponds to the \ghi\ correlation in the 
case of a wind density profile (Nava et al. 2006). 
This relation is shown in Fig. \ref{fg0} with the solid black line (labelled \ep$\propto$\egamma). 
The simulated \G\ distribute \ep\ between 1.5 keV (\G=1) and $\sim$20 MeV (\G=8000).  
 
\item We assign to each simulated burst a jet opening angle \th$\in[1^{\circ}, 90^{\circ}]$ 
extracted from a specified distribution.

\item The probability for a burst to be observed from the Earth depends on the 
viewing angle \thv\ between the jet axis and the line of sight of the observer. 
We extract randomly a viewing angle \thv\ from the cumulative distribution of the  
probability density function $\sin$\thv. 

\item In order to compare the simulated bursts with the source count distribution of existing 
samples of GRBs (see \S3) we compute the observer frame peak fluxes $P$ and fluences $F$. 
To this aim we assume a typical spectrum described by the Band function (Band et al. 1993), 
with low and high photon spectral indexes $\alpha=-1.0$  and $\beta=-2.3$, respectively (i.e. 
corresponding to the typical values observed by different instruments -- e.g. Kaneko et al. 2006; 
Sakamoto et al. 2011)\footnote{These values are also assumed by S12 to constrain the LF of GRBs.}.  
The fluence ${F}$ of each simulated burst in a given energy range is computed by re-normalizing 
this spectrum through the bolometric fluence ${F_{\rm bol}}$=\eiso$(1+z)/4\pi d_{\rm L}^2$, 
where $ d_{\rm L}^2$ is the luminosity distance for a given redshift $z$. 
To derive the peak flux $P$, we assign to each burst an (observer frame) duration $T_{90}$ 
extracted from a distribution centered at 27.5 s and with a dispersion $\sigma_{\rm Log T_{90}}=0.35$. 
This distribution is truncated at $T_{90}=2$ s because we consider only
long duration GRBs in this analysis. 
Such a duration distribution is similar to that of the \fe/GBM GRBs (Paciesas et al. 2012; 
Goldstein et al. 2012) and  includes also very long bursts with $T_{90}\sim$300 s.  
We  assume that the bursts have a simple triangular  light curve and derive the peak luminosity 
as $L_{\rm peak}=2$\eiso$(1+z)/T_{90}$. The peak flux $P$ in a given energy range is obtained 
by re--normalizing the spectrum through the bolometric peak flux $P_{\rm bol}=L_{\rm peak}/4\pi d_{\rm L}^2$. 
\end{enumerate}

\subsection{Computation of \eiso}

The isotropic equivalent energy \eiso\ of  the simulated bursts can be derived from \egamma. 
Since \th$\le$90$^\circ$, simulated bursts cannot be in region II of Fig. \ref{fg0} and \eiso\ 
can  take values on the right hand side of the limit of Eq.~\ref{lghi} shown in Fig. \ref{fg0}. 
According to the values of \th\ and \G\ assigned to each simulated bursts, the isotropic 
equivalent energy is: 
\begin{equation}
E_{\rm iso}=E_{\gamma}/(1-\cos\theta_{\rm jet}) \,\,\,\,\, 
{\rm if} \,\,\,\,\,\, 1/\Gamma_{0}\leq \sin \theta_{\rm jet}
\label{eq1}
\end{equation}
\begin{equation}
E_{\rm iso}=E_{\gamma}(1+\beta_{0})\Gamma_{0}^{2} \,\,\,\,\,\,\,\,\,\,\,\,\,
{\rm if}\,\,\,\,\,\, 1/\Gamma_{0}> \sin \theta_{\rm jet}
\label{eq2}
\end{equation} 
In the latter case \eiso\ is smaller than in  Eq. \ref{eq1}. 
This introduces a limit in the \ama\ plane  of Fig. \ref{fg0} corresponding to the line: 
\begin{equation}
E_{\rm peak} \propto \left[\frac{E_{\rm iso}}{(5-2\beta_{0})^{3}(1+\beta_{0})}\right]^{1/3}
\label{gammalimit}
\end{equation}
(labelled \ep$\propto$\eiso$^{1/3}$ in Fig. \ref{fg0}). For a given \th, bursts with 
a small value of \G\ will have an \eiso\ computed through Eq. \ref{eq2} and will lie on 
the limiting line of region III in Fig. \ref{fg0}. Their radiation is, indeed, 
collimated within an angle $\arcsin$(1/\G) which is larger than their \th.

Simulated bursts can populate the region delimited by boundaries (I, II and III) in Fig.~\ref{fg0}.
This is one (among others)  observational constraint that we will adopt in our simulations (\S3)
to constrain the distributions of \G\ and \th\ and their possible relation. 
According to the relative values of \th, \G\ and \thv, simulated bursts 
are classified as: bursts ``pointing to us" (PO, hereafter), i.e. those  
that can be seen from the Earth, with $\sin$\thv$\le$max[$\sin$\th, 1/\G] and 
bursts pointing in other directions (NPO, hereafter), i.e. not observable from the 
Earth, with $\sin$\thv$>$max[$\sin$\th, 1/\G]. 
We will compare the PO simulated bursts 
with our observational constraints, while the entire population of simulated bursts 
(i.e. PO and NPO) will be used to infer the properties of GRBs (e.g. the distributions 
of \G\ and \th\ and the true burst rate).

\section{Observational constraints}

In order to test whether \th\ and \G\ assume characteristic values or not we   
compare the population of simulated bursts with real samples of GRBs. 
In this section we describe our observational constraints. 
We consider the ensemble of GRBs detected by the Burst Alert Telescope (BAT) on board \sw, 
the Gamma Burst Monitor (GBM) on board \fe\ and the Burst And Transient Source Experiment (BATSE) 
on board the {\it Compton Gamma Ray Observatory} (\cgro). 

\subsection{The \sw\ BAT complete sample}

Salvaterra et al. (2012 - S12 hereafter) constructed a  sample of bright \sw\ bursts consisting 
of 58 GRBs detected by \sw/BAT with $P\ge P_{\rm lim}$=2.6 ph cm$^{-2}$ s$^{-1}$ 
(integrated in the 15--150 keV energy range). Fifty four of these events have a measured redshift $z$ so that the 
S12 sample is 90\% complete in redshift.  Forty six (out of 54) GRBs in this sample have well determined spectral 
properties (filled circles in Fig. \ref{fg0}) and define a statistically robust \ama\ correlation with rank correlation coefficient 
$\rho=0.76$ and chance probability $P=7\times10^{-10}$ (Nava et al. 2012, N12 
hereafter)\footnote{The 8 GRBs without a secure estimate of the redshift or with incomplete 
spectral informations are consistent with the \ama\ correlation defined by the 46 GRBs 
discussed here, see N12 for details.}. 
The correlation properties (slope and normalization) 
of the complete \sw\ sample are consistent with those defined with the incomplete larger sample of 136 
bursts with known $z$ and spectral parameters (see N12). 
Therefore, the distribution of the \sw\  complete sample (46/54 events with well constrained 
\ep) in the \ama\ plane is representative of the larger (heterogeneous) population of 
GRBs with measured $z$ and well constrained spectral properties. 
The 46 GRBs of the complete \sw\ sample define a correlation 
$E_{\rm p}\propto E_{\rm iso}^{0.61\pm0.06}$ 
(shown by the dot--dashed line in Fig. \ref{fg0}) with a scatter (computed perpendicular 
to the best fit line) with a Gaussian dispersion $\sigma=$0.29 dex.

The \sw\ complete sample of S12 contains $\sim$1/3 of the bursts detected by 
\sw\footnote{http://swift.gsfc.nasa.gov/docs/swift/archive/grb\_table/}  
with $P\ge$2.6 ph cm$^{-2}$ s$^{-1}$. 
We verified that the \sw\ complete sample of 54 events selected by S12 is representative 
of the larger population of 149 long \sw\ bursts with $P\ge$\flim: the Kolmogorov--Smirnov 
test on the peak flux distribution of the two samples gives a probability of 0.6 that the two 
distributions are drawn from the same parent population. These bursts were not included in the 
selection of S12 because they do not have favorable conditions for ground--based follow up.

These 149 events with $P\ge$\flim\ are the bursts detected by \sw\ in $\sim$7 yrs from 
its launch within the (half coded) field of view of $\sim$1.4 sr of BAT.  
This corresponds to an average \sw\ detection rate of $\mathcal{R_{\rm Swift}}\sim$15 events sr$^{-1}$ yr$^{-1}$.

\subsection{The \fe\ GBM sample}

Another observational constraint that we consider is the population of bursts detected by 
the GBM on board \fe. 
The spectral properties of GBM bursts have been studied in Nava et al. (2011a) and compared to those of 
BATSE bursts in Nava et al. (2011b). 
More recently, the first release of the GBM spectral catalog (Goldstein et al. 2012) provided 
the spectral parameters and derived quantities (i.e. peak fluxes and fluences) 
for 487 GRBs detected by the GBM in its first 2 years of activity.  
398 bursts in this catalog are long events and have measured peak flux $P$ 
and fluence $F$ (both integrated in the 10 keV--1 MeV energy range)\footnote{$P$ and $F$ are reported in  
Goldstein et al. (2012) and were obtained by integrating the model that best fits the peak time resolved spectrum 
and the time averaged spectrum, respectively.}. 

We cut the GBM sample to $P\ge P_{\rm lim}=2.5$ ph cm$^{-2}$ s$^{-1}$, in order to account for the possible incompleteness of the sample at lower fluxes, obtaining 312 GBM bursts.

The GBM is an all sky monitor that observes on average $\sim$60--70\% of the sky. 
Therefore, the average GBM detection rate is $\mathcal{R_{\rm GBM}}\sim$21 events sr$^{-1}$ yr$^{-1}$   
with peak flux, integrated in the 10 keV--1 MeV energy range, $P\ge 2.5$ ph cm$^{-2}$ s$^{-1}$.

\subsection{The \cgro\ \ba\ sample}

We also consider  the sample of GRBs detected by  \ba. 
The 4B sample (Meegan et al. 1998)  contains 1540 long events and 1496 of these have their 
$P$ and $F$ (both integrated in the 50--300 keV energy range) measured. 
The sample of 1496 \ba\ bursts is cut at  $P\ge P_{\rm lim} = 1$ ph cm$^{-2}$ s$^{-1}$
with 716 \ba\ bursts above this threshold.
Considering the average portion of the sky observed by \ba,  i.e. $\sim$70\% of the sky, 
the detection rate of \ba\ is $\mathcal{R_{\rm BATSE}}\sim$16 events sr$^{-1}$ yr$^{-1}$ for 
GRBs with a peak flux, integrated in the 50--300 keV energy range, 
$P\ge$ 1 ph cm$^{-2}$ s$^{-1}$.

The lower detection rate of \ba\ with respect to GBM is due to the different energy range where 
the peak fluxes are calculated (i.e. 10 keV--1 MeV for GBM and 50--300 keV for \ba, respectively). 
We verified that by considering the GBM bursts with peak flux $P$
integrated in the same energy range of \ba\ (i.e. 50--300 keV) larger than 1 ph cm$^{-2}$ s$^{-1}$ 
(i.e. the same threshold adopted for \ba), the GBM rate is equal to the \ba\ one.

\subsection{Extraction of results}

From each simulation we extract three populations of GRBs among the bursts pointing to us (PO):
\begin{enumerate}

\item the \sw\ comparison sample:  {\it simulated} GRBs with peak flux, integrated 
in the 15--150 keV band, larger than 2.6 ph cm$^{-2}$ s$^{-1}$. 
We also require that their observer frame peak energy is in the range 15 keV--2 MeV. 
Indeed, this is the energy range where \epo\ can be measured by presently flying satellites 
like \sw, Konus and \fe. 

\item the GBM comparison sample: {\it simulated} bursts with a peak flux, integrated in 
the 10 keV--1 MeV energy range, larger than 2.5 ph cm$^{-2}$ s$^{-1}$;

\item the \ba\ comparison sample: {\it simulated} bursts with a peak flux, 
integrated in the 50--300 keV energy range, larger than 1.0 ph cm$^{-2}$ s$^{-1}$.
\end{enumerate}

The simulation is adjusted so that the \sw\ comparison sample contains 149 GRBs, 
i.e. the same number of bright bursts detected by \sw\ (\S3.1). 
Therefore, the \sw\ rate $\mathcal{R_{\rm Swift}}$ is imposed. 
What we derive instead from the simulation is the rate of GBM and \ba\ GRBs that we  compare 
with the real rates of these two instruments described in \S3.2 and \S3.3 respectively. 

We also require that the \sw\ comparison sample is consistent with the \sw\ complete sample of S12. 
To this aim we compare them in the rest frame \ama\ plane and in the observer frame 
\epof\ plane deriving a 2 dimensional Kolmogorov--Smirnov (KS) probability (one for the \ama\ 
and one for \epof\ plane). 
We also verify through a 1 dimensional KS test that the redshift distribution 
of the \sw\ comparison sample is consistent with that of the \sw\ complete sample. 
Finally we compare, through a 1D--KS test, the fluence and peak flux distributions of 
the GBM and \ba\ comparison samples with those of the real samples of GRBs detected by these 
instruments and described in \S3.2 and \S3.3, respectively.

Since the \sw\ complete sample contains only the brightest \sw\ bursts it maps the high $P$ end of the 
peak flux distribution of GRBs. The  GBM and \ba\ samples that we adopt here extend the comparison sample 
to lower values of $P$ and ensures that our simulations  reproduce also the 
faint end of the GRB population\footnote{The $P$ values of the GBM sample are computed on the broad 10 keV--1 MeV energy range (i.e. much broader than the 15--150 keV energy range of \sw). This ensures that the selected sample of the GBM bursts extends the population of GRBs to lower fluxes than those of the \sw\ complete sample.}.

For each simulation we derive the following probabilities:
\begin{itemize}

\item the 2D-KS probability that the \sw\ comparison sample is consistent with the  
complete \sw\ sample  of S12 in the \ama\ plane;

\item the 2D-KS probability that the \sw\ comparison sample  is consistent with the  
complete \sw\ sample of S12 in the \epof\ plane;

\item the 1D-KS probability that the \sw\ comparison sample has a redshift distribution 
consistent with that of the S12 \sw\ sample;

\item the 1D-KS probabilities that the GBM comparison sample is consistent with the GBM 
sample in terms of peak flux $P$ and fluence $F$;

\item the 1D-KS probabilities that the \ba\ comparison sample  is consistent with the \ba\ 
sample in terms of peak flux $P$ and fluence $F$;

\item we verify if the GBM rate predicted by the simulation is consistent, at 1$\sigma$, 
with the GBM rate $\mathcal{R_{\rm GBM}}$.

\item we verify if the BATSE rate predicted by the simulation is consistent, at 1$\sigma$, 
with the \ba\ rate $\mathcal{R_{\rm BATSE}}$.

\end{itemize}
For the KS probabilities we set a limit of 10$^{-3}$ below which we consider that two 
distributions (either 1D or 2D) are inconsistent at more than 3$\sigma$. 
Each simulation, with its assumptions on the distribution  of \th\ and \G,  
is repeated 1000 times and we compute the percentage $\mathcal{P}$ of repeated 
simulations that produce  GRB samples (i.e. \sw, GBM and \ba\ comparison samples) 
consistent with our observational constraints.  

\begin{figure*}
\begin{center}$
\begin{array}{cc}
\hskip -1.5truecm\includegraphics[width=9cm,trim=50 20 40 40,clip=true]{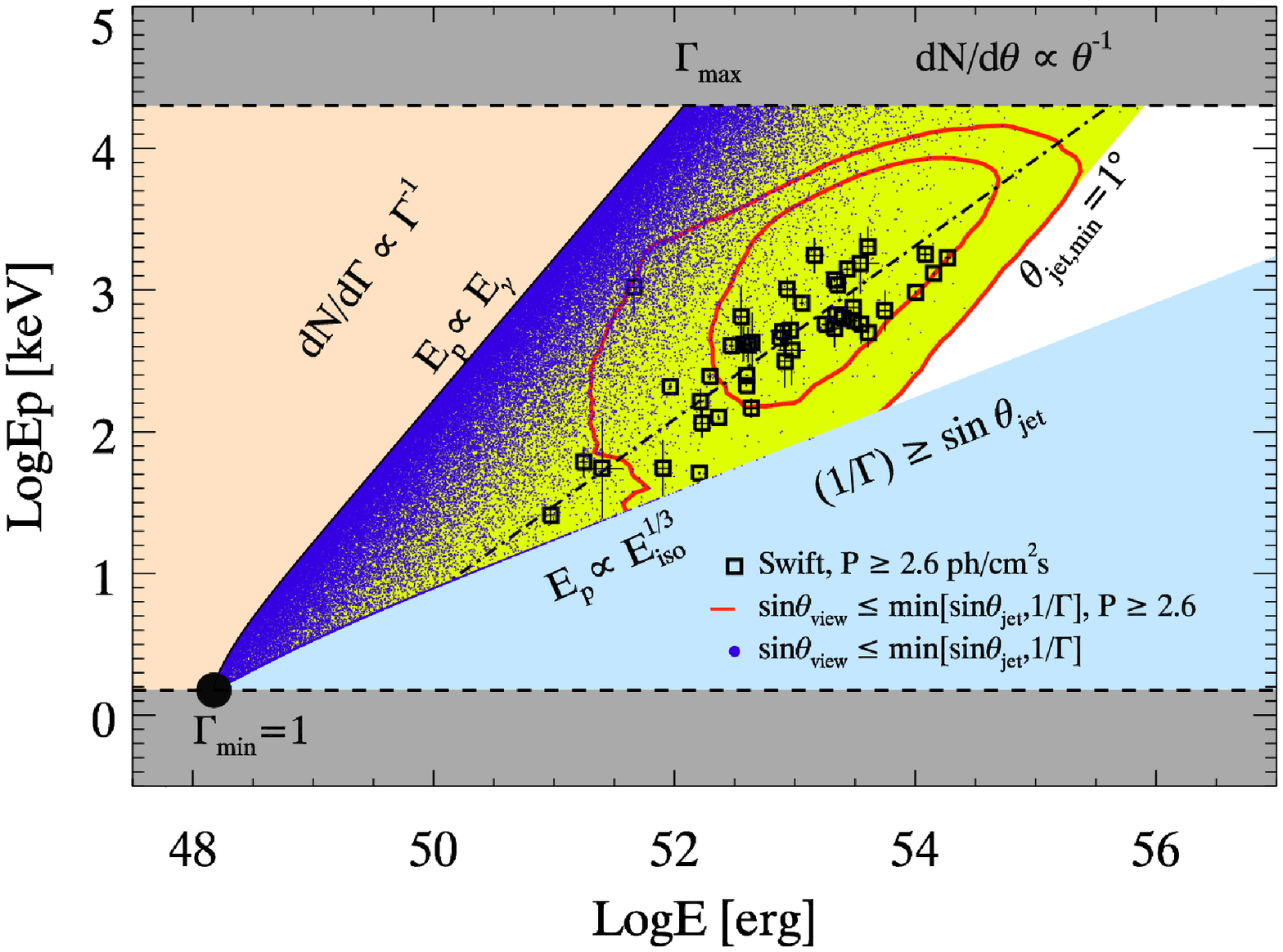} & 
\hskip -0.4truecm\includegraphics[width=9cm,trim=50 20 40 40,clip=true]{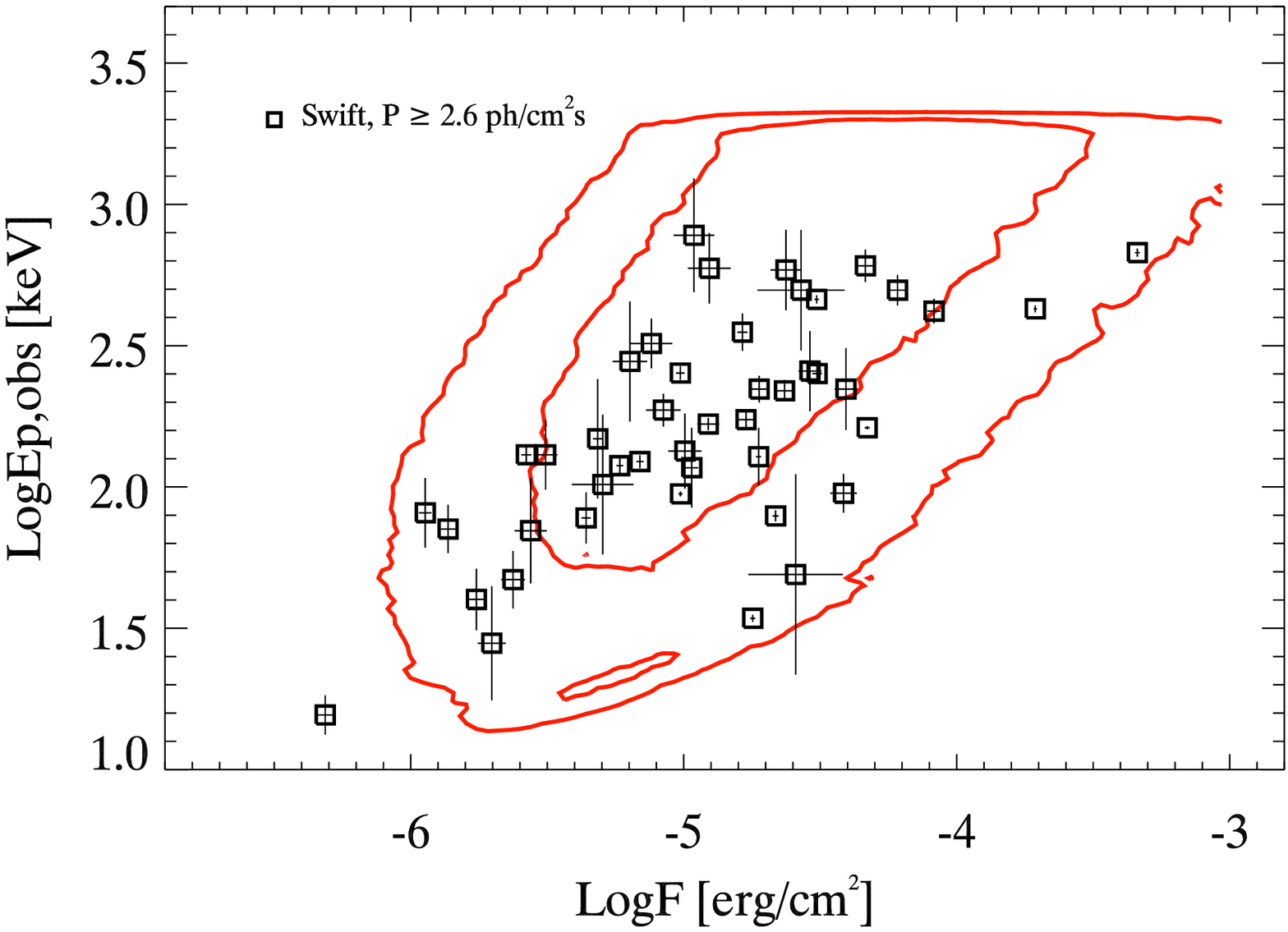} \\
\hskip -1.5truecm\includegraphics[width=9cm,trim=20 10 20 20,clip=true]{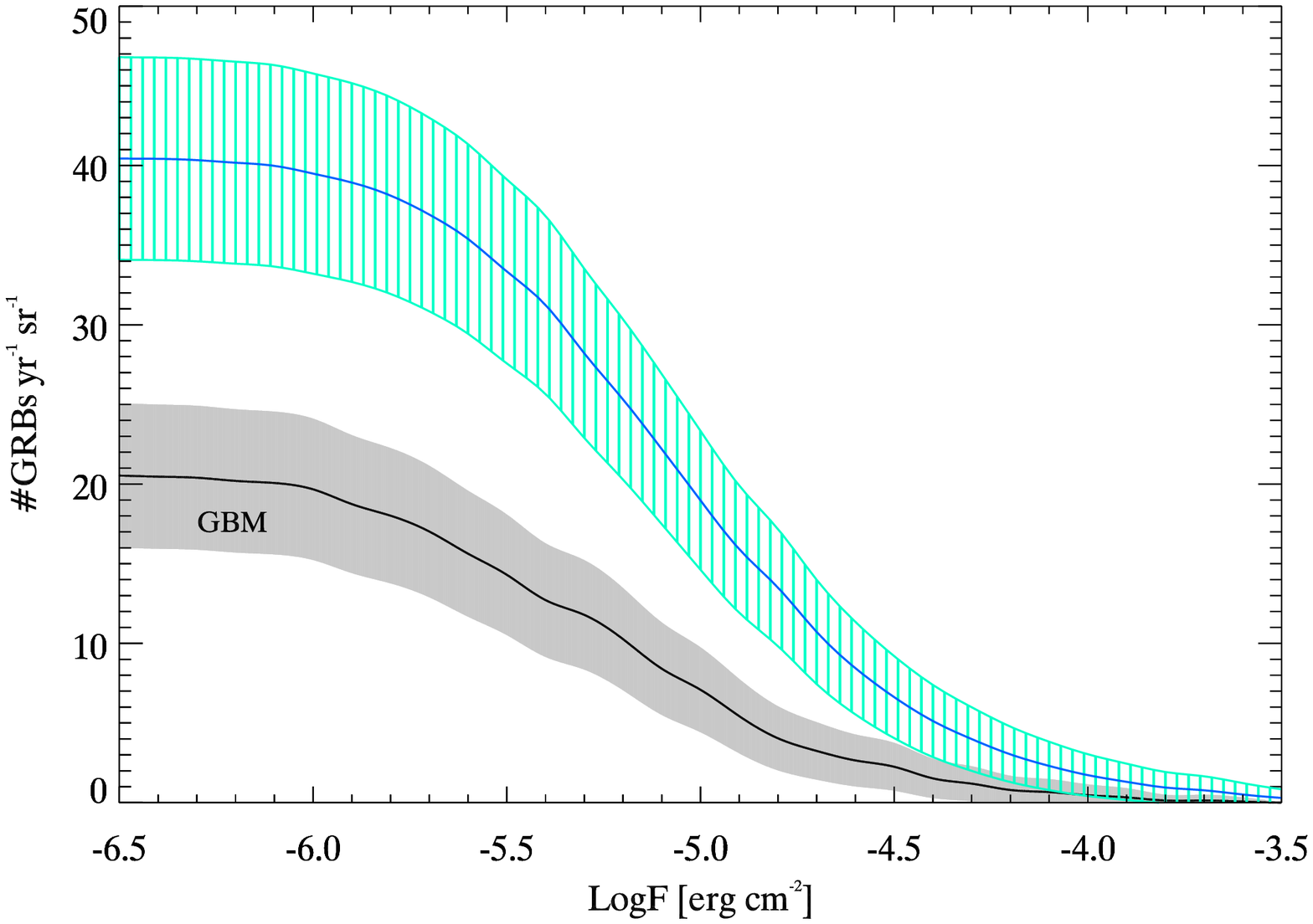} & 
\hskip -0.4truecm\includegraphics[width=9cm,trim=20 10 20 20,clip=true]{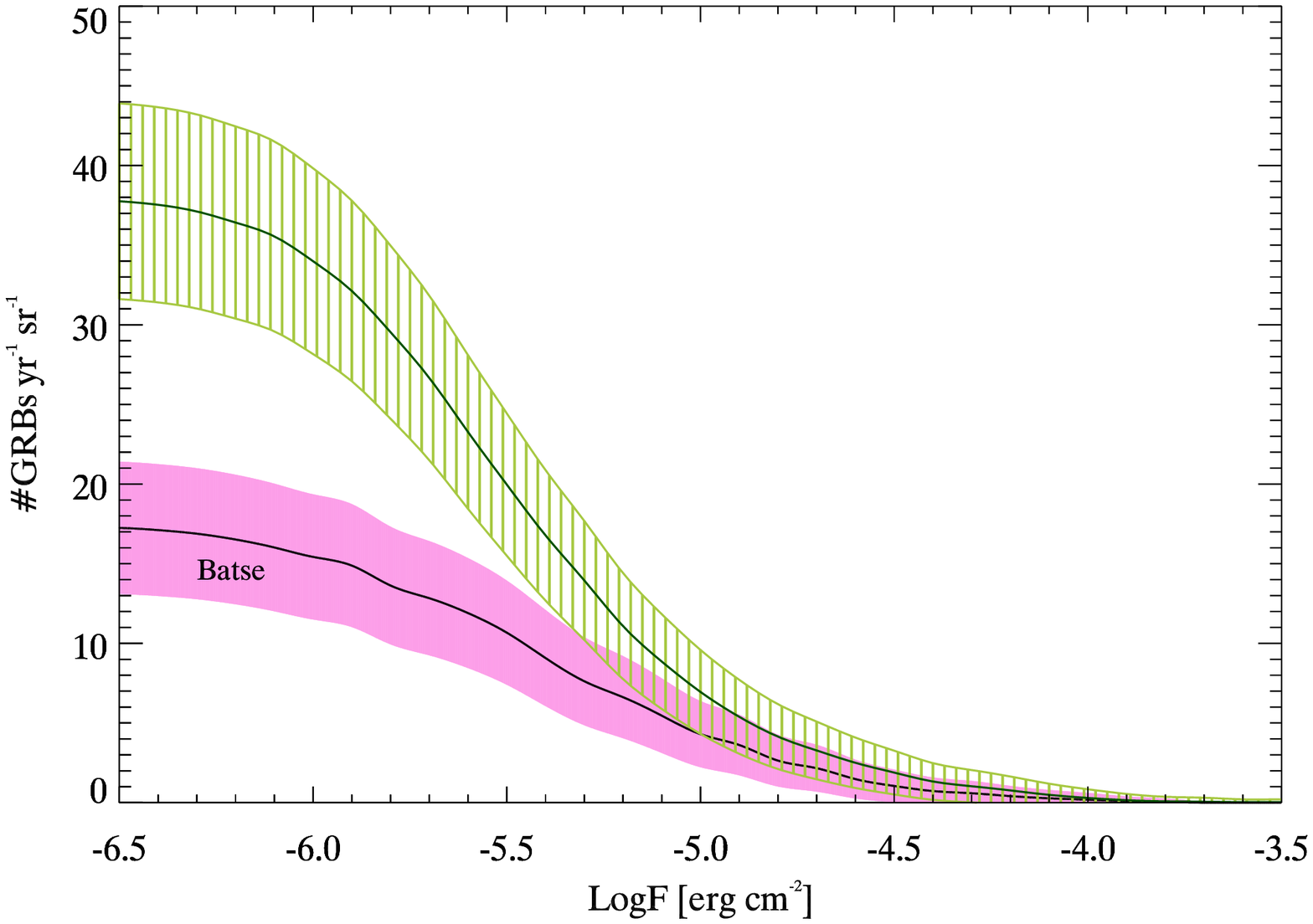} \\
\end{array}$
\end{center}
\caption{
Result of the simulation 
assuming a power law distributions of \th\ and \G. 
{\it Top left}: rest frame \ama\ plane (regions and labels as in Fig \ref{fg0}). 
The simulation assumes $a=c=-1.0$ (see \S4.1) which correspond to uniform distributions 
in the logarithm of \th\ and \G. 
The yellow dots represent all the simulated bursts, the blue dots show the PO bursts, 
i.e. those pointing to us. 
The red solid lines are the smoothed density contour (1 and 2 $\sigma$ confidence levels) of the 
simulated
\sw\ comparison sample. 
They should reproduce the distribution of the 
real \sw\ GRBs of the complete sample (open black squares). 
{\it Top right}: observer frame \epof\ plane where the 
simulated \sw\ bursts (solid contours) are compared with the 
real GRBs of the \sw\ complete sample 
(open square symbols). 
{\it Bottom left}: cumulative rate distribution of GBM real bursts 
(solid black line) and its 1$\sigma$ uncertainty (grey solid filled region) compared 
with the cumulative rate distribution of simulated GBM bursts 
(solid line and dashed cyan region corresponding to its 1$\sigma$ uncertainty). 
{\it Bottom right}: cumulative rate distribution of \ba\ real bursts (solid black line) 
and its 1$\sigma$ uncertainty (pink filled solid region) compared with the prediction 
of the simulation (dashed region). 
}
\label{fg1}
\end{figure*}

\section{Results}


In the following sections we present the results obtained with 
different possible assumptions for the distributions of \G\ and \th.  
We want to test which one among the possible intrinsic distributions of
\G\ and \th\ that one can think of (e.g. power laws, broken power laws or log--normal) 
best reproduces the observational constraints described in the previous section.

\subsection{Power law distributions of \G\ and \th}

\begin{figure*}
\begin{center}$
\begin{array}{cc}
\hskip -1.5truecm\includegraphics[width=8.5cm,trim=50 20 40 40,clip=true]{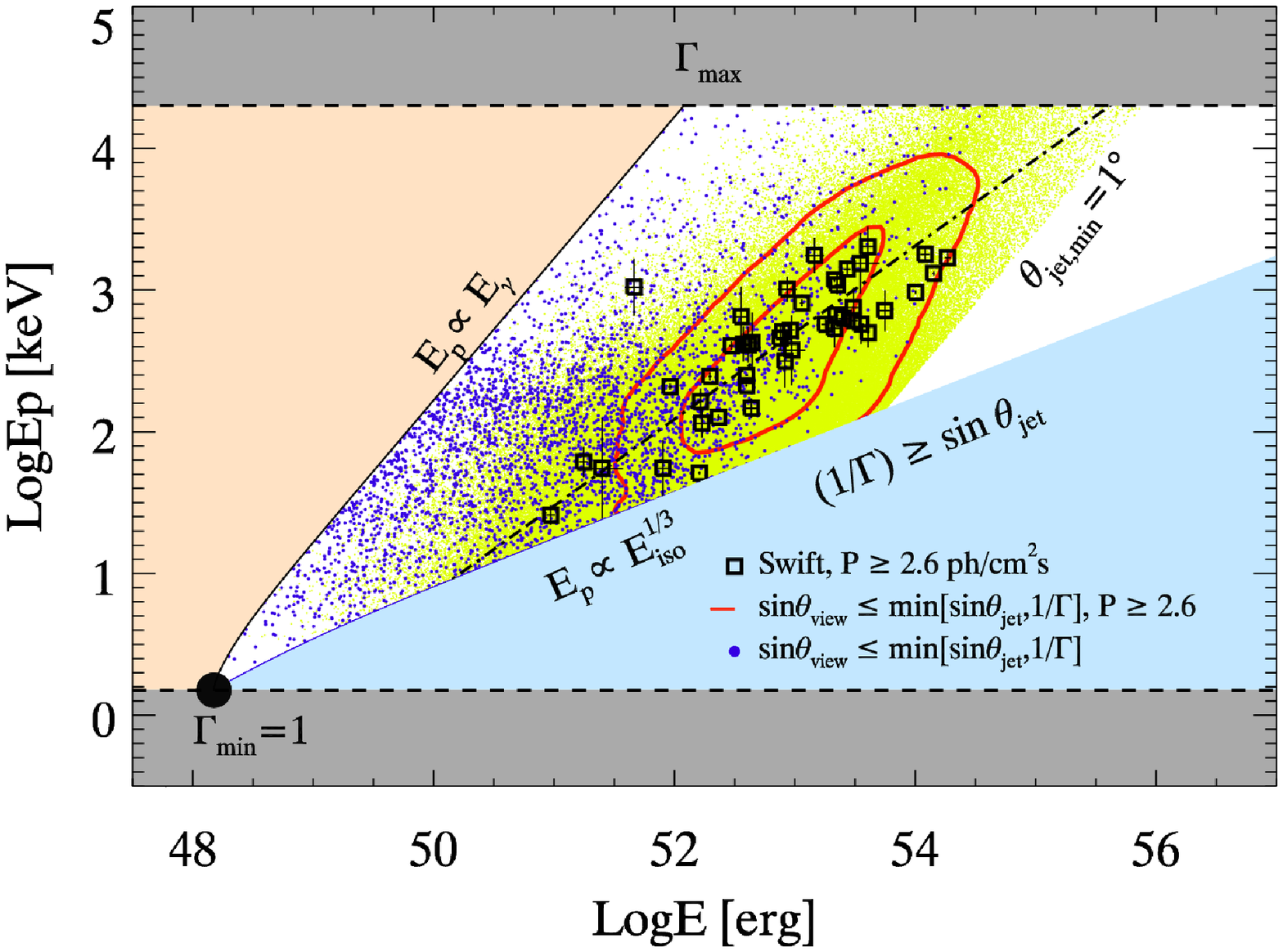} & 
\hskip -0.3truecm\includegraphics[width=8.5cm,trim=50 20 40 40,clip=true]{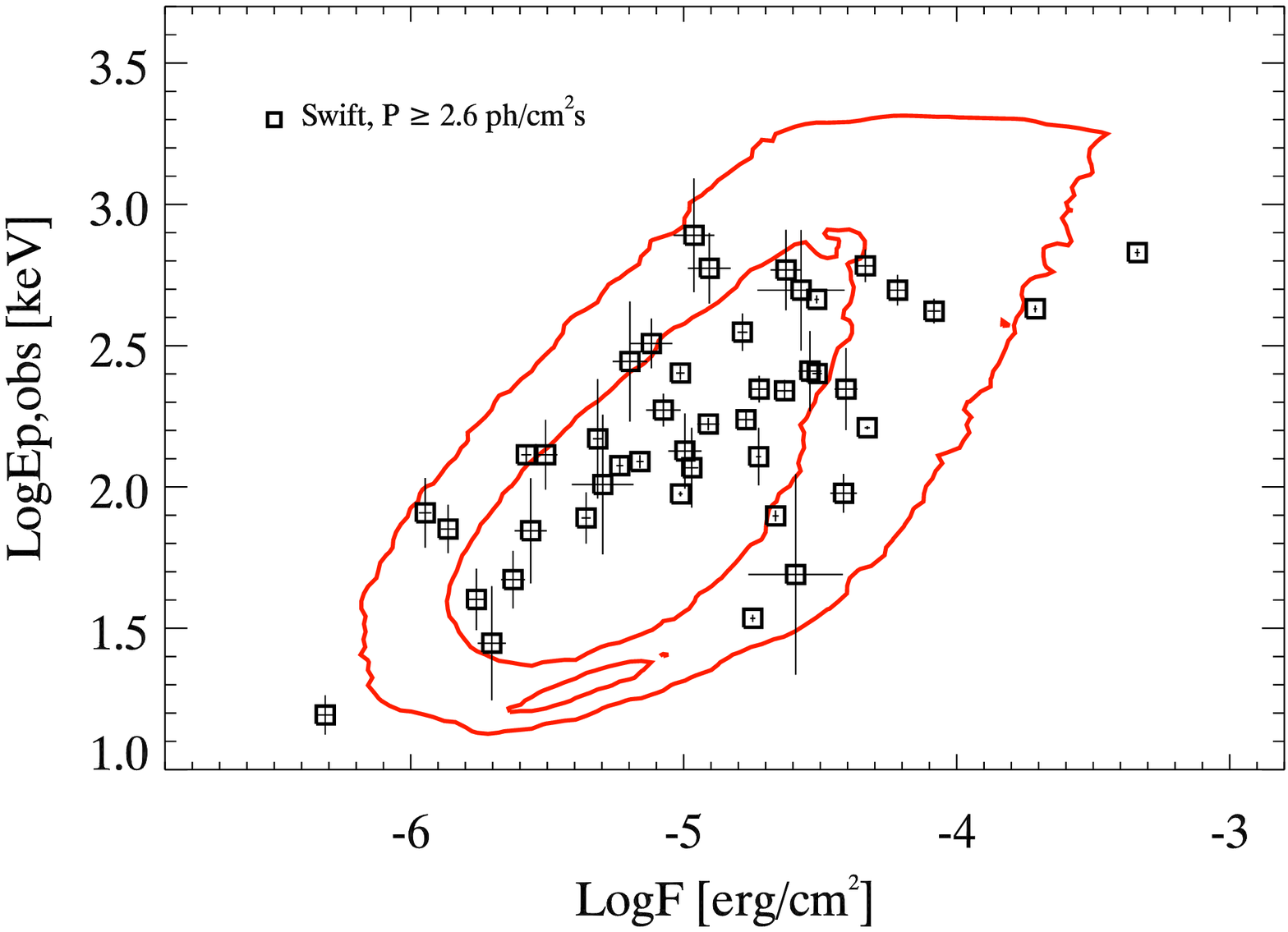} \\
\hskip -1.5truecm\includegraphics[width=8.5cm,trim=20 10 20 20,clip=true]{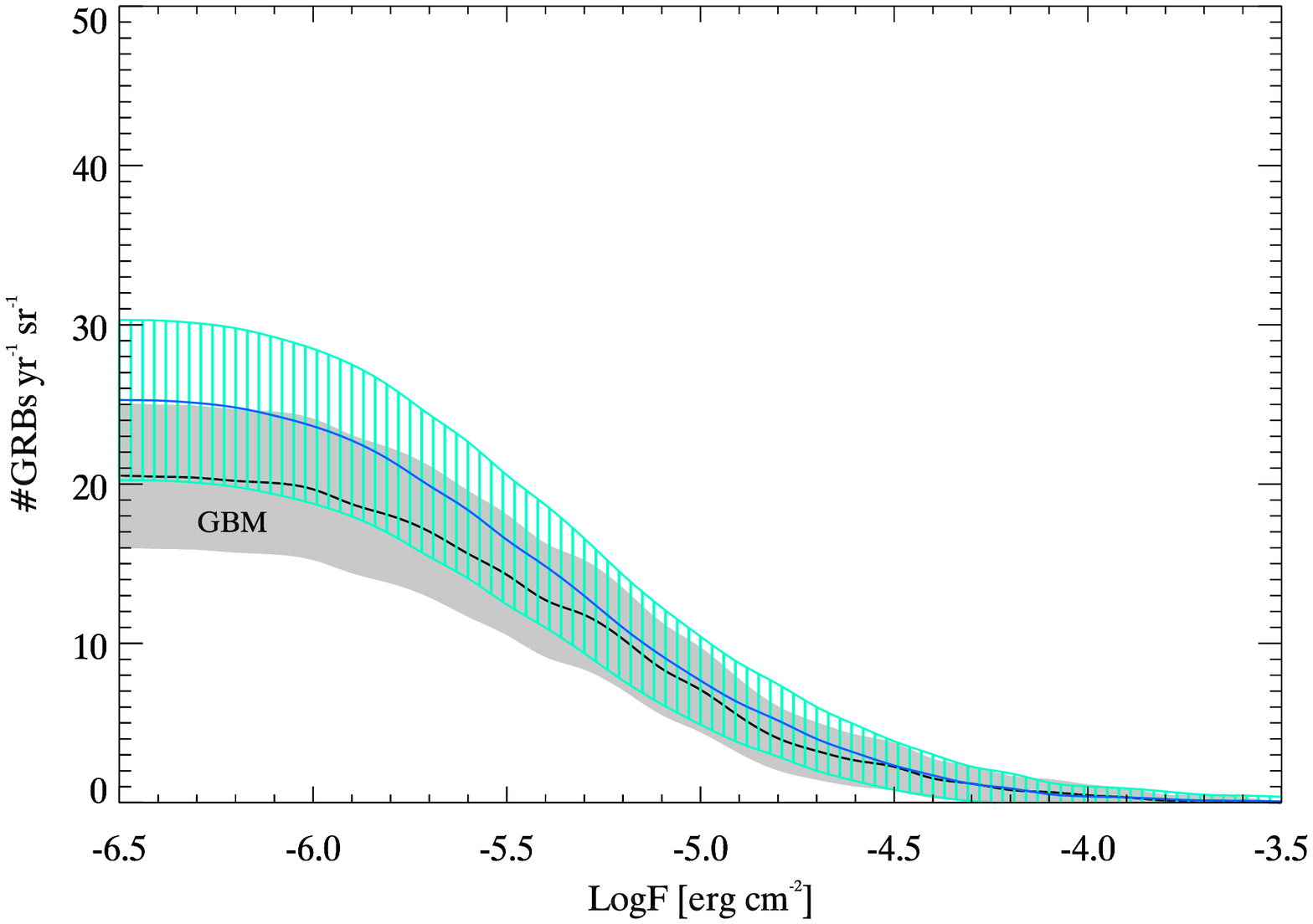} & 
\hskip -0.3truecm\includegraphics[width=8.5cm,trim=20 10 20 20,clip=true]{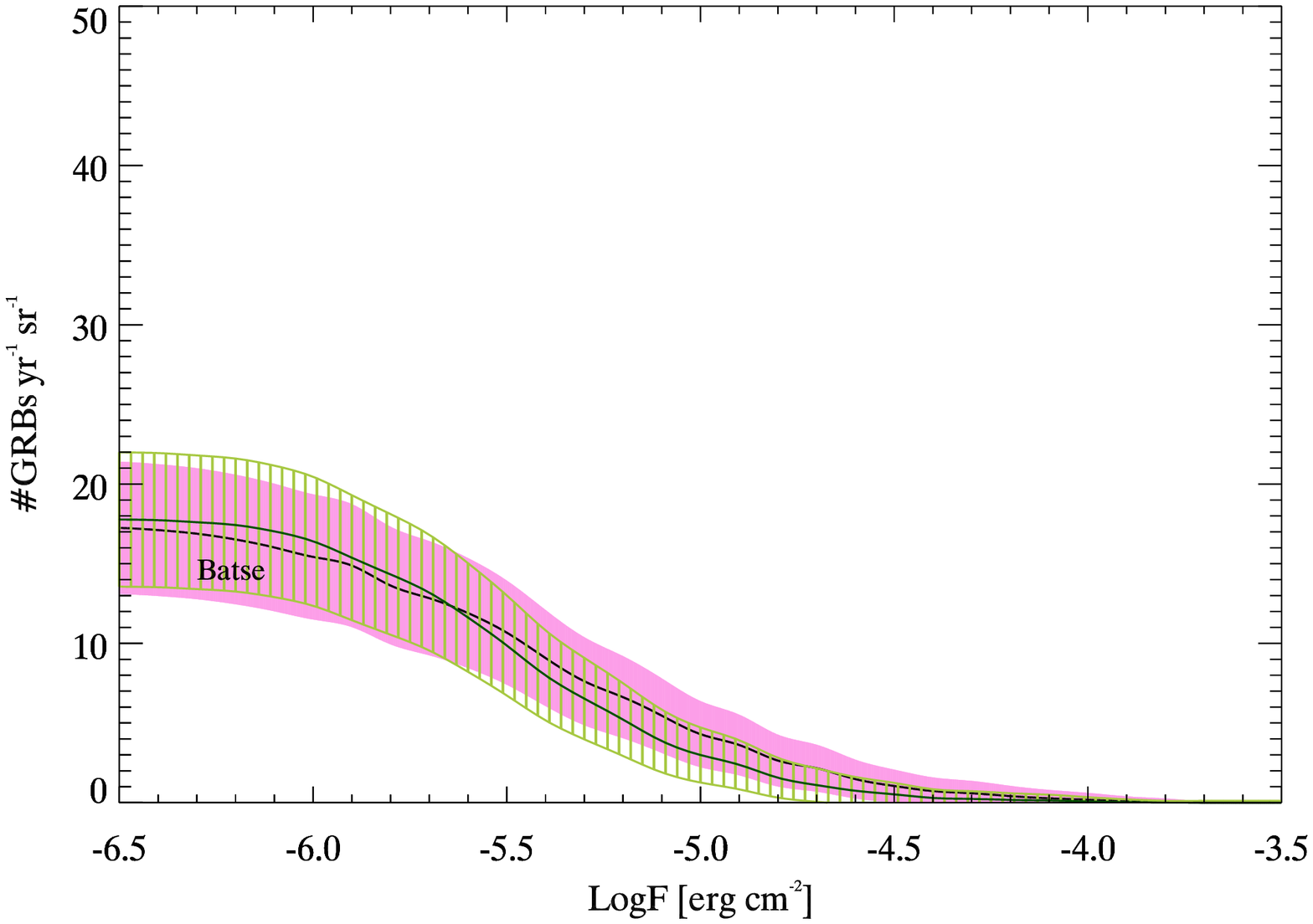} \\
\end{array}$
\end{center}
\caption{
Result of the simulation assuming two
broken power law distributions of \th\ and \G\ (see text for the assumed values of 
the distribution parameters). 
Same plots and symbols as in Fig. \ref{fg1}.
}
\label{fg2}
\end{figure*}

We assume that both \th\ and \G\ are distributed as power laws: 
$dN/d\theta_{\rm jet}\propto \theta_{\rm jet}^{a}$ and $dN/d\Gamma_{0}\propto \Gamma_{0}^{c}$. 
This corresponds to the hypothesis that \th\ and \G\  do not have a characteristic value.
We consider $a\in[-2,-1]$ and $c\in[-2,-1]$. 
 
The choice of these parameters corresponds to have most of the simulated bursts with low \G\ factors and 
with small \th\ values. One could think that such distributions are already excluded by the observed distributions of 
\th\ and \G\ (which are log--normal) discussed in \S1. However, those are the {\it observed} distributions of \G\ and \th\ and 
they are subject to several biases (see \S1). The intrinsic distributions might well be completely different and this motivates to start with this simplest assumption, i.e. that both \G\ and \th\ have power law distributions.

Under the hypothesis that both \G\ and \th\ have power law distributions (with free parameters $a$ and $c$ varied 
in the above ranges with a step 0.2 in both parameters), 
only in 1\% of 1000 repeated simulations  
we can find an agreement with all our observational constraints. 
In order to show the inconsistency of the simulations with the observational constraints we 
present in 
Fig. \ref{fg1}  the  results of the simulations assuming that \th\ and \G\ have power 
law distributions with $a=c=-1$. 
This case, shown as an example, corresponds to a uniform distribution of Log\G\ and Log\th. 
 
The rest frame \ama\ plane (top left panel in Fig. \ref{fg1}) is filled uniformly with 
simulated bursts (yellow dots) distributed between the \ghi\ limit and with a minimum 
\th=1$^\circ$ (the oblique right limit to the distribution of yellow dots). 
The simulated GRBs pointing to us (PO) have preferentially large \th\ values (blue dots in 
the top left panel of Fig.~\ref{fg1}). The simulated bursts of the \sw\ comparison sample 
(here represented by the smoothed density contours\footnote{These are obtained by staking 1000 
simulations and smoothing the obtained distribution in the \ama\ plane.} 
-- red solid lines in Fig. \ref{fg1}) are inconsistent with the real GRBs of the \sw\ complete 
sample (open squares). The red contours extend at high \ep\ values where there is a deficit 
of \sw\ bursts and they also over predict the number of bursts on the right hand side of the 
distribution of the real \sw\ bursts (i.e towards large values of \eiso\ for 
intermediate/high values of \ep).
 
Also in the observer frame \epof\ plane (top right panel in Fig. \ref{fg1}) the simulated 
\sw\ comparison sample (solid contours) are inconsistent with the real \sw\ bursts 
of the complete sample (open squares).  
Simulated bursts of the \sw\ comparison sample tend to concentrate towards the 
upper part of the \epof\ plane. 

The bottom panels of Fig. \ref{fg1} show the cumulative rate distribution of the fluence 
for the GBM and \ba\ sample (right and left panels of Fig. \ref{fg1}) compared with 
the predictions of the simulations (dashed regions in the bottom panels of Fig. \ref{fg1}). 
The rate of GBM and \ba\ bursts predicted by the simulation which assumes a power 
law distribution for both \th\ and \G\ (with index --1) is a factor $\sim$2 
larger than the  rate of GBM bursts. Also the distributions of the peak flux of the simulated \ba\ and GBM 
samples are inconsistent with the real samples. 

If we assume steeper power law distributions of \th\ and \G\ [e.g. $(a,c)=(-2,-2)$], 
the excess of bursts with large peak energy (both in the rest frame and in the observer 
frame of Fig. \ref{fg1}, top left and right panels respectively) is reduced but the rate 
of simulated GBM and BATSE bursts increases becoming more inconsistent with the real rates 
of GRBs detected by these two instruments (bottom panels of Fig. \ref{fg1}). 
This result shows that all the constraints that we have adopted (\S3) are relevant: the GBM and \ba\ comparison sample map 
the low end of the peak flux/fluence distribution while the \sw\ complete sample maps the bright burst tail of such distributions. 
The bursts of the \sw\ complete sample, having their $z$ measured, map the distribution of GRBs in the rest frame \ama\ plane.

\subsection{Peaked distributions of \G\ and \th}


\begin{figure*}
\begin{center}$
\begin{array}{cc}%
\hskip -1.5truecm\includegraphics[width=8.5cm,trim=50 20 40 40,clip=true]{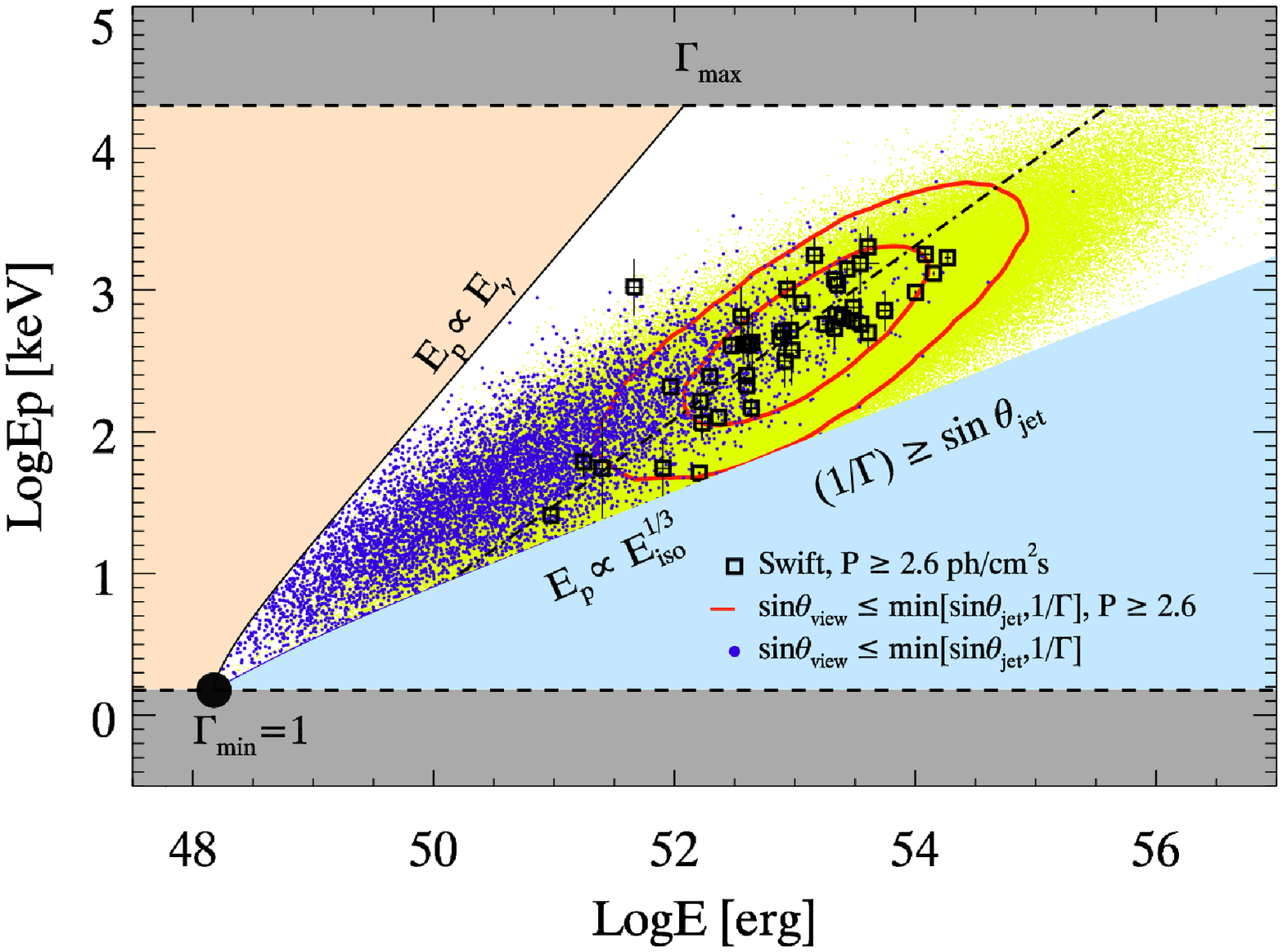} & 
\hskip -0.3truecm\includegraphics[width=8.5cm,trim=50 20 40 40,clip=true]{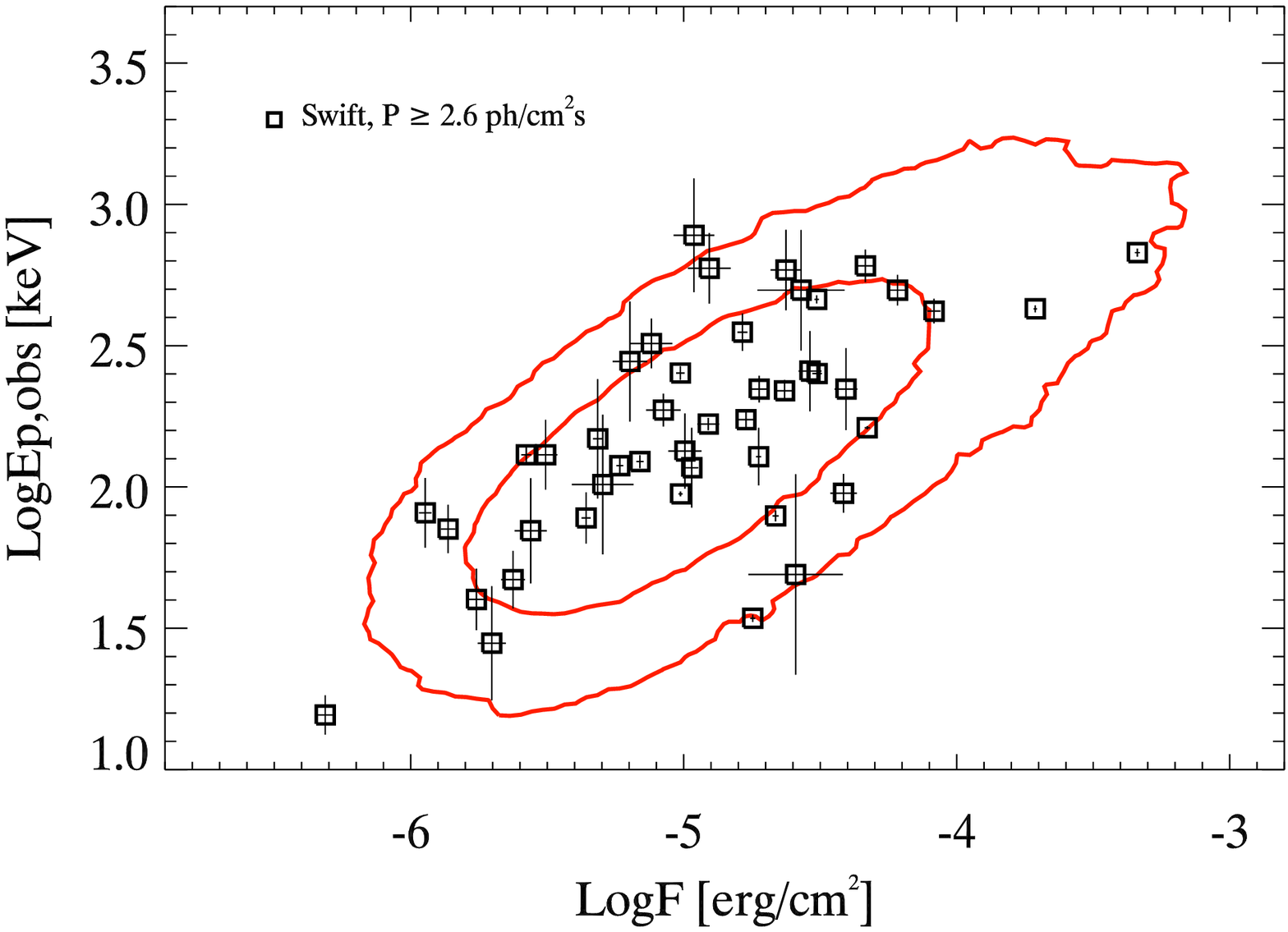} \\
\hskip -1.5truecm\includegraphics[width=8.5cm,trim=20 10 20 20,clip=true]{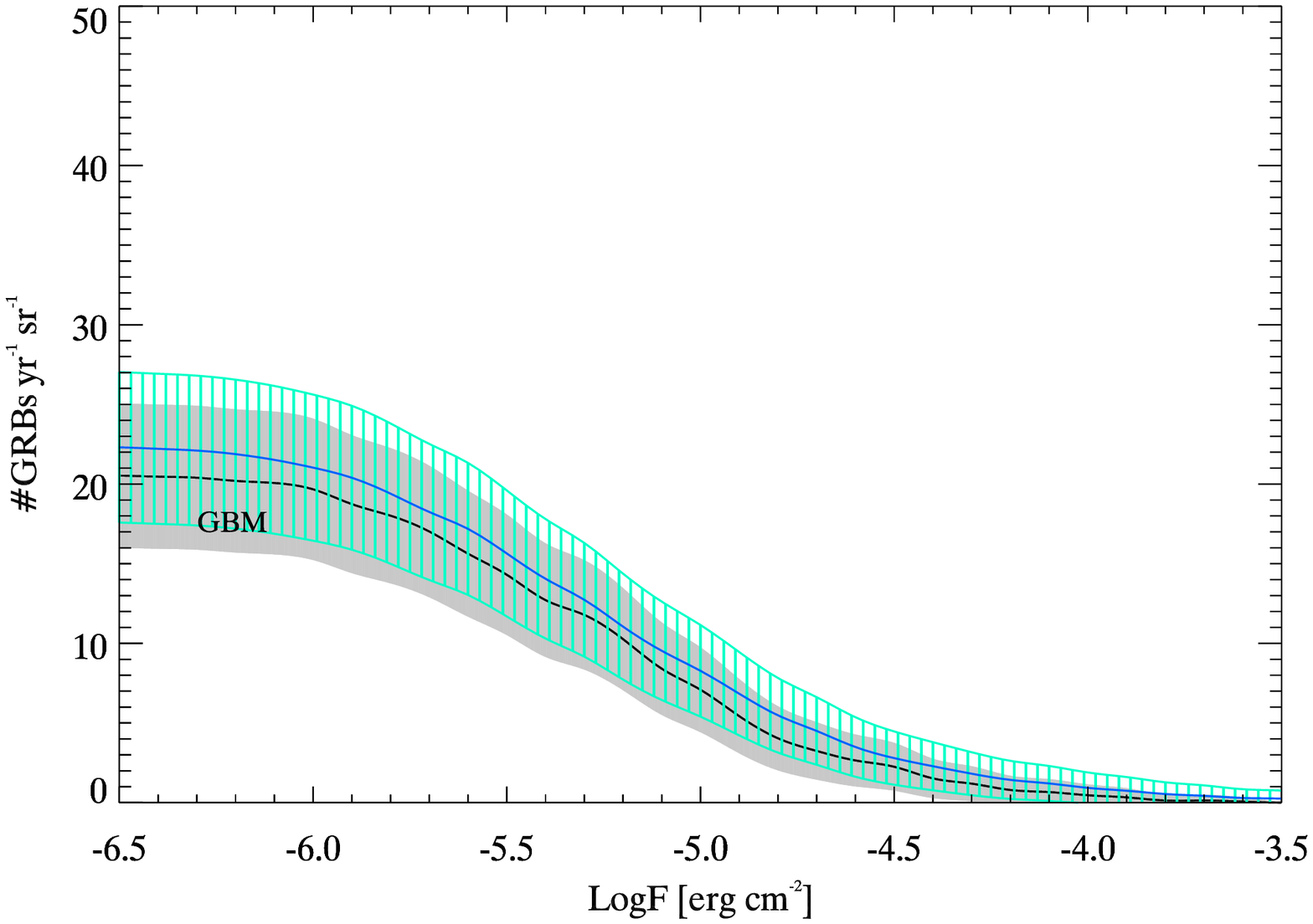} & 
\hskip -0.3truecm\includegraphics[width=8.5cm,trim=20 10 20 20,clip=true]{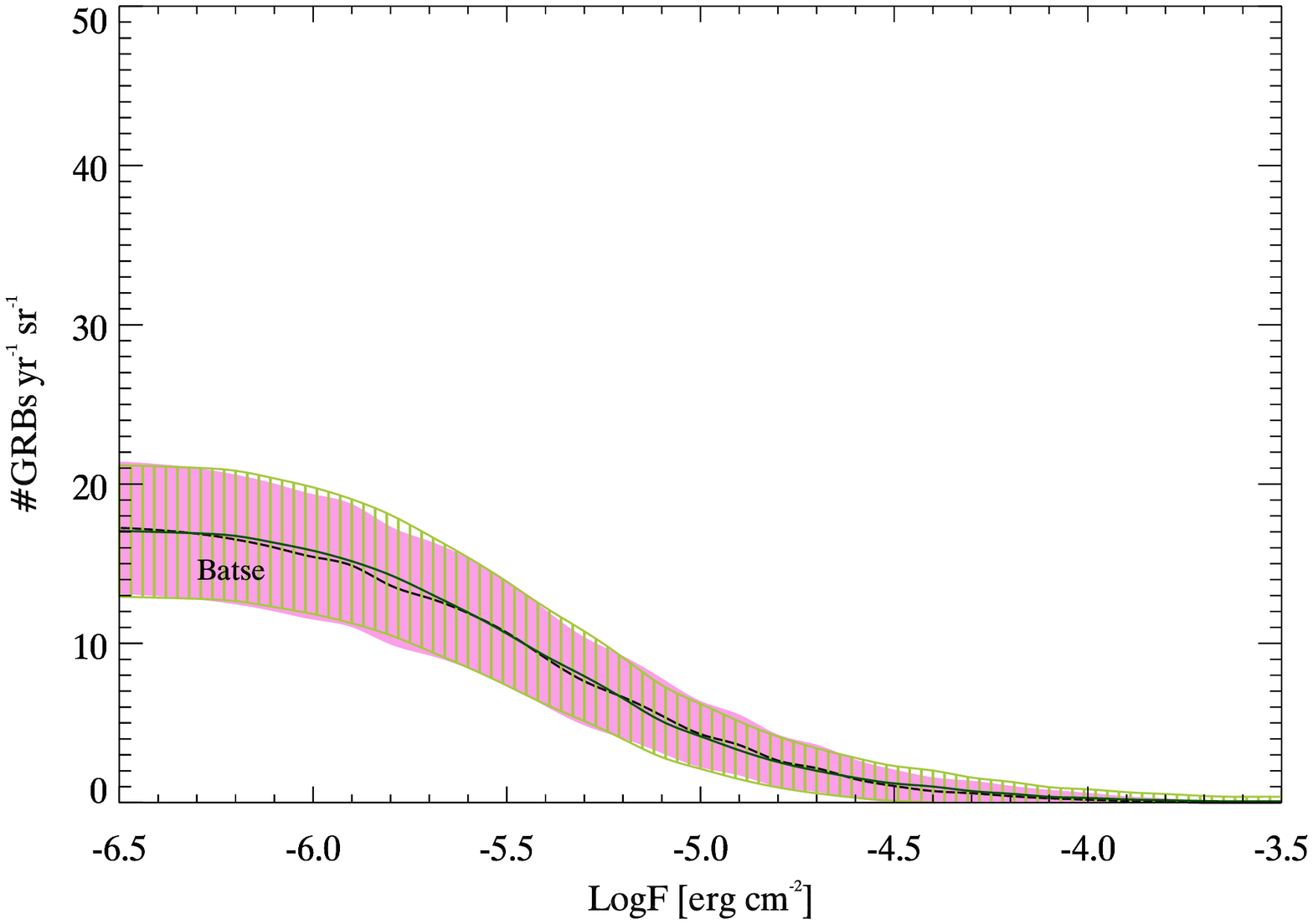} \\
\end{array}$
\end{center}
\caption{Simulations assuming log--normal distributions of \th\ and \G\ and a 
relation between them (see text). Symbols and labels as in Fig. \ref{fg1}.
}
\label{fg3}
\end{figure*}

Since we could not find agreement between the simulations which assume power 
law distributions of \th\ and \G\ and our observational constraints, we now 
consider the case of peaked distributions of \th\ and \G.

The simplest assumption is that  \th\  and/or  \G\ are distributed as broken power laws. 
We first assumed that only \th\ or \G\ have a broken power law distribution, while the 
other parameter is distributed as a single power law. 
In this case we cannot find a percentage of repeated simulations larger than 2\% in 
agreement with our observational constraints.
  
We then considered the case of a broken power law distribution for both \th\ and \G:
$$
\frac{dN}{d\theta_{\rm jet}} = \left\{ \begin{array}{rl}
 \theta_{\rm jet}^{a} &\mbox{ if \,\, $\theta_{\rm jet}\le \theta_{*}$} \\
 \theta_{\rm jet}^{b} &\mbox{ if \,\, $\theta_{\rm jet}> \theta_{*}$}
       \end{array} \right.
$$
$$
\frac{dN}{d\Gamma_{0}} = \left\{ \begin{array}{rl}
 \Gamma_{0}^{c} &\mbox{ if \,\, $\Gamma_{0}\le \Gamma_{*}$} \\
 \Gamma_{0}^{d} &\mbox{ if \,\, $\Gamma_{0}> \Gamma_{*}$}
       \end{array} \right.
$$
For the distribution of \th\ we consider the following parameter ranges: 
$a\in[0.5, 2.0]$, $b\in[-2.0, -5.0]$ and $\theta_{*}\in[3^{\circ}, 12^{\circ}]$. 
For \G:  $c\in[0.5, 2.0]$, $d\in[-2.0, -5.0]$ and $\Gamma_{*}\in[50, 120]$.  
The free parameters  are varied with 
step 0.1 for $a$ and $b$ and 0.5$^\circ$ for $\theta_{*}$ for the broken power law 
distribution of \th\ and with step 0.1 for $c$ and $d$ and 10 for $\Gamma_{*}$ 
for the broken power law distribution of \G.

We find that at most $\sim$20\% of the 1000 repeated simulations reproduce our observational 
constraints when $(a,b,\theta_{*})=(0.5,-3.0,4.5^{\circ})$ and $(c,d,\Gamma_{*})=(1.8,-3.5,70)$ 
with step 0.1 for $c$ and $d$ and 10 for $\Gamma_{*}$ for the broken power law distribution of \G.  
A lower percentage of agreement is obtained for any other choice of the free parameters.

We show in Fig. \ref{fg2} the results of the simulations with the above parameter 
values for the distributions of \th\ and \G. 
We note that a better agreement is now found between the rate of the GBM and \ba\ bursts 
(bottom panels of Fig. \ref{fg2}) while the distribution of simulated bursts of the \sw\ 
comparison sample (solid contours) are inconsistent with the \sw\ bursts of the complete 
sample both in the rest frame \ama\ plane (top left panel of Fig. \ref{fg2}) and in the 
observer frame plane \epof\ (top right panel of Fig. \ref{fg2}).  

The assumption of a characteristic value of \G\ corresponds to concentrate GRBs around a 
typical value of \ep\ (see Eq. 2). 
In this case the narrower \th\ distribution 
reduces the number of simulated bursts with
large values of \th, thus clustering the simulated GRBs  
of the PO class
around the \ghi\ limit of Fig. \ref{fg2} (top left panel) 
that was found in the case of single power laws (\S4.1).
 
A broken power law is a simple approximation of a peaked distribution. 
The real distribution of \G\ and \th\ could have a different shape. 
We then considered the case of log--normal distributions for both \G\ and \th, 
with central values of the \th\ distribution between 3$^\circ$ and 12$^\circ$ (step 0.5$^\circ$) 
and   width between 0.3 and 0.8 (step 0.05) and central values of \G\ between 50 and 120 (step 5)
and width between 0.2 and 0.8 (step 0.05). 
We find that, if \th\ has a log--normal 
distribution with a median value of 4.5$^\circ$ (with a dispersion of 0.5) and \G\ 
is distributed as a log--normal with median 85 (with a dispersion of 0.45), the 40\% 
of the 1000 repeated simulations is in agreement with all our observational constraints. 

The latter assumption, that seems to improve the consistency between the simulated GRB population and the
observational constraints, suggests that \G\ and \th\ have log--normal distributions. However, the fact that  no more than 40\% 
of the repeated simulations can reproduce all our observational constraints, is suggesting that some ingredient is 
still missing. This is the subject of the next section where we study for the first time through our numerical simulations, the
possibility that there is a relation between the average values of \th\ and \G.

\subsection{The relation between \th\ and \G}

By assuming a  \th\ distribution with a characteristic value as in \S4.2, 
the simulated bursts in the \ama\ plane cluster around a 
correlation which is linear in this plane (i.e. parallel to the \ghi\ limit), while the 
\ama\ correlation defined by the \sw\ complete sample (and similarly by the larger, 
incomplete, sample of bursts with measured redshift -- see N12) has a flatter 
slope, i.e. \ep$\propto E_{\rm iso}^{0.6}$. 
In other words, for an infinitely narrow distribution of \th, the simulated bursts 
(yellow dots in Fig. \ref{fg2} top left panel) would  produce a linear \ama\ correlation 
which is inconsistent with the observed \ama\ correlation. 
This suggests that, besides the fact that  \th\ and \G\ should have characteristic values
(i.e. peaked distributions), {\it they should also be correlated}.

Indeed, G12 find that the comoving frame properties of GRBs (and in particular 
the fact that \ep$\propto$\G\ and \eiso$\propto$\G$^2$) can be combined to explain 
both the \ep$\propto E_{\rm iso}^{0.5}$ and the \ep$\propto$\egamma\ correlation if 
the ansatz \th$^{2}$\G=const is valid.  
Several recent numerical simulations of jet acceleration in GRBs suggest that a link between 
\G\ and \th\ should exist, although the form of this relation  depends on 
several assumptions of these simulations.
In this section we explore, for the first time, if a relation \th$^{m}$\G=$K$  can account for the 
observational constraints described in \S3 and in this case we constrain its free parameters ($m$ and $K$).
We start from the result of the previous section, which showed that the best result (i.e. 40\% of the repeated 
simulations are in agreement with the observations) is obtained assuming two log--normal distributions 
for \G\ and \th.

We simulate bursts with  Log\G\ distributed as a Gaussians with a characteristic 
central value Log$\Gamma_{*}$ and a dispersion $\sigma_{\rm Log\Gamma_{0}}$. 
Similarly we assume a Gaussian distribution for Log\th\ centered at Log$\theta_{*,\rm jet}$ 
and with a dispersion $\sigma_{\rm  Log\theta_{\rm jet}}$. 
We then assume that there is a relation between \th\ and \G\ of the form 
Log$\theta_{*,\rm jet} = -1/m$Log$\Gamma_{*} + q$. 
In this way the distribution of Log\th\ is centered on a value which is 
given by the assumed relation between \th\ and \G. 

We explored the parameter space (defined by 5 free parameters) and found that 
80\% of our simulations are consistent with our constraints if Log$\Gamma_{*}=1.95$ 
with a dispersion of $\sigma_{\rm Log\Gamma_{0}}=0.65$ dex, $m=2.5$, $q=1.45$ and 
$\sigma_{\rm Log\theta_{\rm jet}}=0.3$ dex.

We show in Fig. \ref{fg3} the results of this simulation which assumes 
log--normal distributions of \G\ and \th\ and a relation between these two parameters. 
In the \ama\ plane (top left panel in Fig. \ref{fg3}) and in the \epof\ plane 
(top right in Fig.~\ref{fg3}) we find a good agreement between the simulated 
\sw\ comparison sample (solid contours) and the real \sw\ complete sample (open squares).
Now the predicted rate of GBM and \ba\ bursts is fully consistent with the 
real ones (bottom left and right panels in Fig. \ref{fg3} respectively).  

We stress that, given the assumptions of our simulation (e.g. the  spectrum, duration 
and unique values of the comoving frame energetics of all GRBs) we do not expect 
to find 100\% of the simulations reproducing our constraints.  
However, we can use our code to derive interesting properties of the population of GRBs. 
Indeed, in our simulations we generate a population of GRBs pointing in every direction. 
Only those pointing towards the Earth (PO) are then compared with existing
samples of GRBs (like those described in \S3). 
This is also the population of bursts that will be explored by 
future GRB detectors with better sensitivity than the present ones. 
We can derive the properties of the whole GRB population  (i.e. all the bursts pointing in 
whatever direction), like the jet opening angle 
distribution, the bulk Lorentz factor distribution and the true GRB rate.

%
\begin{table*}
\begin{center}
\begin{tabular}{lllllll}
\hline
\hline
Distrib.  &sample   &$\sigma$        &$\mu$	        &Mode           &Mean           & Median      \\
\hline
\th\      &ALL      &0.916$\pm$0.001 &1.742$\pm$0.002 &2.47$^\circ$   &8.68$^\circ$   &5.71$^\circ$  \\
          &PO       &0.874$\pm$0.010 &3.308$\pm$0.013 &12.73$^\circ$  &40.04$^\circ$  &27.33$^\circ$ \\
          &PO*      &0.610$\pm$0.020 &2.83$\pm$0.029  &11.68$^\circ$  &20.41$^\circ$  &16.95$^\circ$ \\
          &PO \sw\  &0.527$\pm$0.032 &1.410$\pm$0.043 &3.10$^\circ$   &4.71$^\circ$   &4.10$^\circ$  \\ 
          &PO* \sw\ &0.544$\pm$0.298 &1.043$\pm$0.434 &2.11$^\circ$   &3.29$^\circ$   &2.83$^\circ$  \\ 
                     
            \\                                        
\G\       &ALL      &1.475$\pm$0.002 &4.525$\pm$0.002 &11    &274     &92     \\
          &PO       &1.452$\pm$0.020 &2.837$\pm$0.025 &2     &49      &17       \\
          &PO \sw\  &0.975$\pm$0.060 &5.398$\pm$0.083 &85    &355     &221     \\

\hline
\end{tabular}
\caption{
Parameter values ($\mu$ and $\sigma$) obtained by fitting a log--normal function (Eq. \ref{lgn}) 
to the distributions of \th\ (Fig. \ref{fg4}, \ref{fg5}) and \G\ (Fig. \ref{fg6}) for 
all the simulated bursts (ALL), for those pointing to us (PO) and for those pointing 
to us and with a peak flux larger than \flim, i.e. the \sw\ comparison sample (PO \sw).  
(*) fit of the distributions of GRBs pointing towards the Earth that should not have a 
jet break (see \S5.2). For each distribution are reported the three moments: the mode, 
the mean and the median.  
}
\label{tab1}
\end{center}
\end{table*}

\subsection{\th\ distribution of GRBs}

\begin{figure}
\hskip -0.5truecm
\includegraphics[width=9cm,trim=20 10 20 20,clip=true]{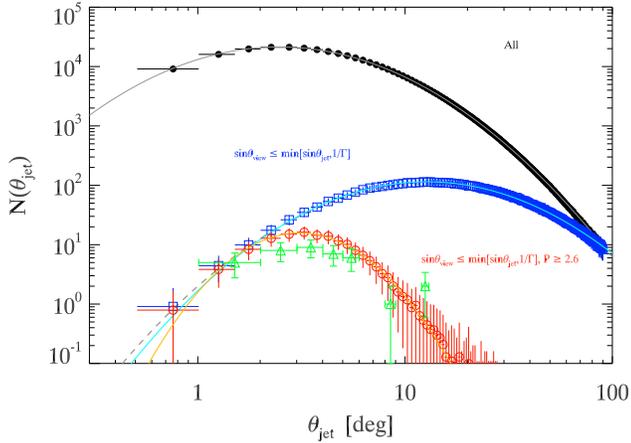}
\caption{Distribution of \th\ of GRBs. The \th\ distribution of the total sample of simulated GRBs is shown by the filled circles. The solid grey line shows the fit with a lognormal function (Eq.\ref{lgn}). The subsample of GRBs pointing towards the Earth (PO) is 
shown by the open (blue) squares and its fit with a lognormal by the cyan line. The sample of PO GRBs and peak flux $P\ge$2.6 cm$^{-2}$ s$^{-1}$ (i.e. the \sw\ comparison sample) is shown by the open (red) circles and its lognormal fit by the orange line. The dashed (grey) line shows the lognormal fit of the distribution of all the bursts (solid grey line) multiplied by $1-\cos\theta_{\rm jet}$. The green triangles show the distribution of the 27 GRBs with measured jet opening angle collected in Ghirlanda et al. (2004, 2007). 
}
\label{fg4}
\end{figure}

From the best simulation described in \S4.3 we can derive the distribution of the 
jet opening angle of GRBs. 
In Fig. \ref{fg4} we show the distribution of \th\ for all the simulated bursts 
(black points) and for the PO bursts (open cyan squares). 
The population of GRBs pointing towards the Earth and with a peak flux 
$P\ge$2.6 cm$^{-2}$ s$^{-1}$ in the 15--150 keV range (i.e. the \sw\ comparison sample) 
is shown by the open (red) circles. 
All the distributions of \th\ can be modeled with a log normal function:
\begin{equation}
N(x)=\frac{A}{x \sigma \sqrt{2\pi}}\exp\left[-\frac{(\ln x-\mu)^{2}}{2\sigma^{2}}\right]
\label{lgn}
\end{equation}
where the free parameters are $(\mu,\sigma)$ and the normalization $A$. 
The best fit parameters $\mu$ and $\sigma$ are reported in Tab. \ref{tab1}. 
The peak of the log--normal distribution, i.e. its mode, is $\exp(\mu-\sigma^2)$, 
the mean is $\exp(\mu+\sigma^{2}/2)$ and the median is $\exp(\mu)$. 
Since the asymmetry of the log--normal distributions can be considerably large, 
we report in Tab. \ref{tab1} all these moments. 

The \th\ of GRBs of the \sw\ comparison sample (red open circles in Fig. \ref{fg4}) 
have a mean of \th$\sim4.7^\circ$. 
This distribution is consistent with the \th\ estimated from the break 
of the optical light curves (Ghirlanda et al. 2004, 2007), 
shown by the open (green) triangles in Fig. \ref{fg4}.

The GRBs that point to the Earth (PO - shown by the open blue squares in Fig.~\ref{fg4}) 
have a \th\ distribution peaking at considerably larger values (40$^\circ$ - see Tab.1) than the
entire GRB population. This can be easily interpreted: consider the distribution of the entire 
population of GRBs (black dots in Fig.~\ref{fg4}) which contains all bursts pointing in every direction. 
The probability that a  burst with a certain \th\ is pointing to us is 
proportional to $(1-\cos \theta_{\rm jet})$. Therefore the distribution of \th\ for PO bursts is obtained from the total distribution 
by multiplying by $(1-\cos \theta_{\rm jet})$. This reduces the number of bursts per unit \th\ and also shifts the peak of the PO distribution 
towards an average larger value. This is shown in Fig. \ref{fg4} by the dashed (grey) line which is obtained by 
multiplying the fit of the distribution of \th\ of the entire GRB population 
(solid gray line in Fig. \ref{fg4}) by  $(1-\cos \theta_{\rm jet})$ and it fits 
the distribution of the PO bursts (open squares in Fig. \ref{fg4}). 

Among the simulated bursts that are pointing towards the Earth we considered the bright bursts (i.e. selected with the same 
peak flux threshold of the \sw\ complete sample). These bursts tend to have small jet opening angles and this accounts for their \th\ distribution peaking at $\sim$5$^\circ$ in Fig.~\ref{fg4} (open red circles).

Although apparently there is a similarity between the \th\ distribution of all bursts (i.e. pointing in every direction) and the \th\ distribution of the PO bright bursts, they differ by a factor 2 (1.8) in their peak values (and dispersions) which are reported in Tab.1. 

The three distributions shown in Fig.~\ref{fg4} allow us to make some further considerations. 
If we could measure \th\ for all bursts pointing towards the Earth (PO in Tab. \ref{tab1}), 
we would obtain the open (blue) square distribution of Fig. \ref{fg4}  with a 
mean $\sim40^\circ$.  However, the real \th\ distribution of the population of GRBs 
(i.e. all the simulated bursts --  black filled circles distribution in Fig. \ref{fg4}) has a mean of $\sim$8.7$^\circ$ 
and it is more consistent with the distribution of the simulated PO bursts with large peak fluxes (the \sw\ comparison sample). 
This suggests that the bursts distributed in the high part of the \ama\ correlation, 
where are the bursts of the complete \sw\ sample (filled black dots in Fig.~\ref{fg3} top left panel),  
properly sample the  peak of the \th\ distribution of the entire GRB population.


\subsection{GRBs with no jet break}
\begin{figure}
\hskip -0.5truecm
\includegraphics[width=9cm,trim=20 10 20 20,clip=true]{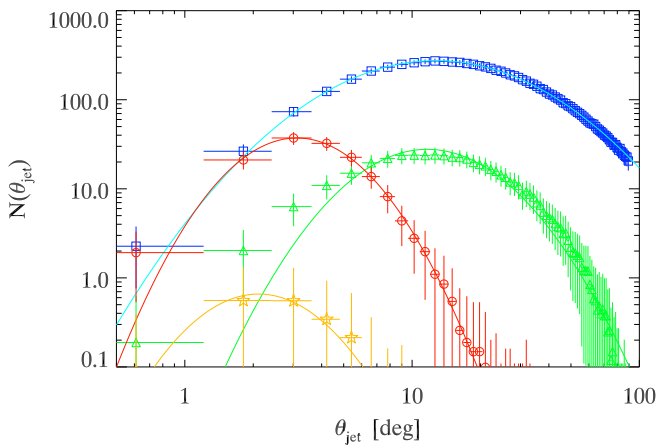}
\caption{
Distribution of \th\ of GRBs: the PO simulated GRBs (open blue squares) and the 
GRBs of the PO class that have $\sin \theta_{\rm jet} \le 1/\Gamma_{0}$ 
(open green triangles) are shown. 
The latter are those that should not show any jet break time in their afterglow light curve. 
For the PO bursts with $P\ge$2.6 cm$^{-2}$ s$^{-1}$ (open red circles) we  show the 
subsample of bursts that have $\sin \theta_{\rm jet} \le 1/\Gamma_{0}$ (open orange stars).
}
\label{fg5}
\end{figure}

It has been shown in \S3 that if a burst has a \G\  such that 
$\sin \theta_{\rm jet} \le 1/\Gamma_{0}$, its \eiso\  is determined by \G\ (Eq. \ref{eq2}) 
and not by \th. 
This value is lower than that computed by \th\ (Eq. \ref{eq1}). 
In these bursts, therefore, we should not observe a jet break in their light curve since 
the emitted radiation is initially collimated within an angle $\arcsin$1/\G\ larger than \th. 
Since $\Gamma$ decreases during the afterglow phase due to the deceleration of the fireball by 
the interstellar medium, in these bursts the jet break, corresponding to the transition 
1/$\Gamma \sim$\th, will never happen. 
 
The above argument contributes to explain the fact that bursts might not show an evident jet break in their 
afterglow light curve if   
1/\G$\ge\sin$\th. However, in these bursts we  expect that the afterglow light curve is declining with a typical post--break decay index $\sim -p$ (where $p$ is the shock--accelerated electron energy distribution index - e.g. Panaitescu \& Kumar 2001).
Other possible explanations for the lack of \tjet\ measurements have been proposed.  
Numerical simulations (e.g. Van Eerten et al. 2010), for instance,  suggest that the jet 
break transition can be very smooth (almost difficult to be distinguished from a single power law decay with available data sets) 
due to a combination of the jet dynamics before and after the jet break time (and additional complications can be induced by the 
viewing angle effects when the observer is not on--axis). Although 
a detailed discussion of the missing jet breaks in GRBs is out of the scope of this paper, we notice that 
bursts with $\sin \theta_{\rm jet} \le 1/\Gamma_{0}$ can partly account for the explanation of the lack of measured 
jet breaks. This is the first time that such an argument is presented and surely deserves further studies.

Fig. \ref{fg5} shows the distribution \th\ of PO bursts (open blue squares) 
and the subsample of bursts with no jet break (open green triangles). 
These amount to $\sim$6\% of PO bursts. 
The mean of their log--normal distribution is \th$\sim$20$^\circ$.  
One testable observational prediction of our simulations is that GRBs with no jet breaks 
should be preferentially soft (\epo\ of few tens of keV) 
The open red circles in Fig. \ref{fg5} correspond to PO bursts of the \sw\ comparison 
sample while the open orange star symbols correspond to  bursts 
with no jet break. 
These have a mean jet opening angle $\sim$3.3$^{\circ}$. 
We find that $\sim$2\% of the \sw\ bright bursts should not have jet break in 
their afterglow light curves. 
They could correspond to those events which do not show any 
evidence of a jet break in their optical light curve (e.g. Mundell et al. 2006; Grupe et al. 2007) 
although other observational selection effects very likely contribute to the 
paucity of the jet break measurements. 
The fit of the distributions shown in Fig. \ref{fg5} with log--normal functions 
are reported in Tab. \ref{tab1}.

\subsection{\G\ distribution of GRBs}

\begin{figure}
\hskip -0.5truecm
\includegraphics[width=9cm,trim=20 10 20 20,clip=true]{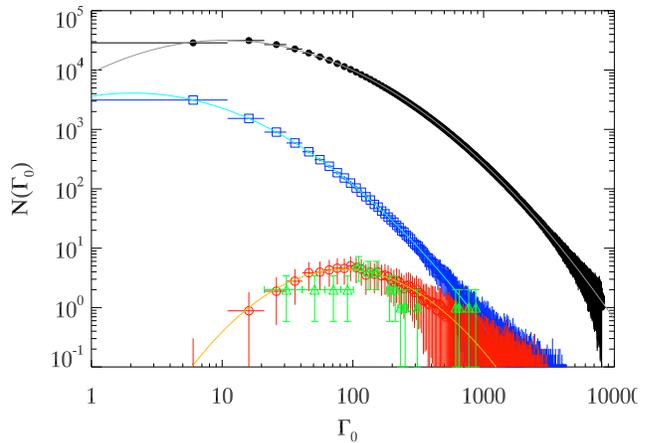}
\caption{
Distribution of \G\ of GRBs. Symbols as in Fig.\ref{fg4}. 
The 30 GRBs with \G\ estimated from the peak of their afterglow light curves 
(G12) are shown with the green open triangles.
}
\label{fg6}
\end{figure}

From our simulation we can derive the distribution of \G\ (Fig. \ref{fg6}). 
The total population of simulated bursts (filled circles in Fig. \ref{fg6}) 
has a log normal distribution with a mean \G=274. 
Those pointing towards the Earth (open blue squares in Fig. \ref{fg6}) have a smaller mean, \G=49. 
The PO bursts with peak flux larger than 2.6 cm$^{-2}$ s$^{-1}$, i.e. those of the \sw\ 
comparison sample, have a typical \G=355. 
Although the distribution of \G\ factors for those bursts with a peak in their 
afterglow light curves (G12) is still made of few events, it agrees (open green triangles 
in Fig. \ref{fg6}) with that predicted by our simulations (for the sample of 
PO bursts of the \sw\ comparison sample -- open red circles in Fig. \ref{fg6}).

Also in the case of \G\ we note that if we were able to measure \G\ for all the bursts that point towards the Earth, 
we would obtain a slightly smaller peak value of \G\ with respect to that of the distribution of all the bursts (pointing in 
every direction.

We note that the \G\ distribution of the general population of GRBs peaks at considerably low values of \G. This is a result of our simulations where, as explained in \S 4.2, we assume a peaked logarithmic distribution of \G\ with free peak and width. If we assume a distribution of \G\ with a smaller fraction of bursts with low \G--values, then we cannot reproduce the flux and fluence distributions and the detection rates of GRB detection of the GBM and \ba\ instruments. Therefore, our simulations 
predict that a considerable fraction of GRBs should have \G\ as low as a few tens. These bursts might well be detected by current instruments. While the detailed study of their prompt and afterglow properties is out of the scope of the present paper, we note that their prompt emission should hardly differ from that of bursts with larger \G\ values (except for the obvious fact that their prompt \ep\ and \liso\ is lower). In fact,  if \G\ is low the fireball deceleration timescale (e.g. Eq.14 in Ghirlanda et al. 2011) is 
$t_{\rm peak}\sim 4\, E_{\rm iso,50}^{1/3} \Gamma_{0,1}^{-8/3}$ hours which is much larger than the prompt emission timescale. So, while the prompt emission of low-\G\ burst should not be influenced by the afterglow contribution, their late time afterglow onset could be a distinctive feature (typical  afterglow onset timescales are of the order of few hundreds second - Ghirlanda et al. 2011).

\subsection{The GRB rate}

\begin{figure}
\hskip -0.5truecm
\includegraphics[width=9cm,trim=20 10 20 20,clip=true]{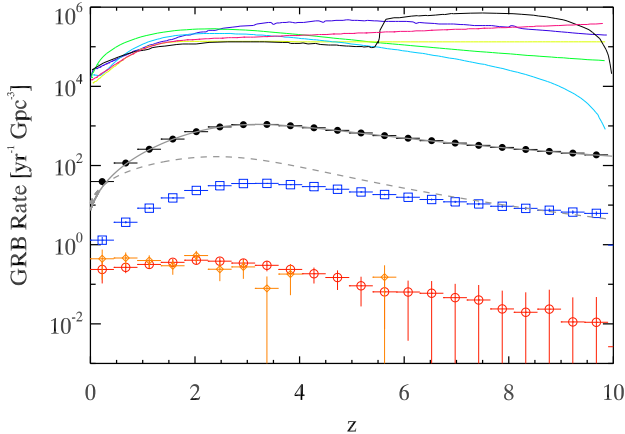}
\caption{
GRB rate as a function of redshift. 
The sample of simulated bursts is shown by the filled
black circles and the GRB formation rate 
assumed in our simulation (see \S2) is shown by the solid grey line. 
This curve is normalized to the histogram. 
The GRB formation rate without density evolution (as derived by Li et al. 2010 -- 
see Eq. \ref{z}) is shown by the  dashed grey line.  
The rate of bursts pointing to us (PO) is shown by the open (blue) squares and 
that of the PO bursts with $P\ge$2.6 cm$^{-2}$ s$^{-1}$ (i.e. the \sw\ comparison 
sample -- PO \sw) by the open (red) circles. For comparison is also shown the rate of 
\sw\ GRBs of the complete sample 
(rescaled to match the rate of PO \sw\ rate).  
The solid lines reported in the top of the plot are the cosmic rates of SNIb/c 
computed by Grieco et al. (2012) for different assumptions on the cosmic star formation rate. 
}
\label{fg7}
\end{figure}

Another consequence of our simulations is the rate of GRBs. 
This is shown as a function of redshift for the entire population of simulated bursts 
(filled circles in Fig. \ref{fg7}), in units of bursts Gpc$^{-3}$ yr$^{-1}$.
The GRB redshift distribution (Eq. \ref{z}) assumed in our simulations is shown by the solid grey 
line in Fig. \ref{fg7} and the observed star formation rate (Li 2008) is shown by the dashed (grey) line rescaled by
an arbitrary factor to match the rate of GRBs at $z=0$. 
We also show the rate of PO bursts (open blue squares) and that of the PO bursts of 
the \sw\ comparison sample (open red circles). 
Fig. \ref{fg7} also shows the recent estimate of the rate of SNIb/c computed by Grieco et al. (2012). 
The different curves for SNIb/c correspond to different assumption of the cosmic star formation 
rate (CSFR) in that paper. 
As a result of our simulation, the local rate of GRBs is $\sim$0.3\% that of SNIb/c.

\section{Summary and discussion}

We have studied two fundamental parameters of GRBs: the jet opening angle \th\ 
and the bulk Lorentz factor \G. 
The first question that we aimed to answer was {\it whether \th\ and \G\ have preferential values}. 
The direct measure of \th\ through the jet break times observed in the optical light curves  
(Frail et al. 2001; Ghirlanda et al. 2004, 2007) shows that \th$\sim$5$^\circ$. 
The measure of \G\ from the peak of the afterglow light curve for $\sim$30 GRBs (G12) 
also shows a characteristic value\footnote{This average value is obtained assuming 
that the circumburst medium has a wind density profile (see G12).} of \G$\sim$60. 
However, the limited number of events with a direct estimate of \th\ and \G\ and 
the possible selection effects, related to the difficulties of measuring these two parameters (see \S1),
prevent us to assume them as representative of the GRB population. 
In particular we want to test the consistency of different possible distributions of \th\ and \G\ 
with a set of available observational constraints (\S2). Moreover, we aim at constraining the 
free parameters of the distributions of \th\ and \G\ and derive if and how these two parameters are
correlated. 

In this paper we used a population synthesis code to simulate GRBs with different 
assigned distributions of \th\ and of \G\ each one with a set of free parameters that we left free to vary within certain ranges.
Obviously, we did not assume the observed distributions of \th\ and \G\ as constraints to avoid
circularity. 

We assume that GRBs have a unique comoving frame peak energy \epcom\ and  
collimation--corrected energy \egcom\ (the large black dot in Fig.~\ref{fg0}) 
which are transformed into their corresponding rest frame \ep\ and \egamma\ respectively.
The assigned \th\ and \G\ allow us to derive the isotropic equivalent energy of the 
simulated bursts according to the relative value of \th\ and \G.  
\eiso$\sim$\egamma$/\theta_{\rm jet}^2$ if $\sin\theta_{\rm jet}>1/\Gamma_0$,
while \eiso$\sim$\egamma$\Gamma_0^2$ in the opposite case. 
This introduces a ``natural bias" in the distribution of \eiso: 
those bursts with a small enough \G\  will have an isotropic energy which 
is smaller than that one would calculate using the value of \th. 
In the \ama\ plane of Fig. \ref{fg0} this corresponds to a limit \ep$\propto E_{\rm iso}^{1/3}$. 
Bursts with 1/\G$>$sin\th\ will lie along this limiting line and there should be no 
GRBs on the right of this line [i.e. in region (III) in Fig. \ref{fg0}].
 
This is the first time that the limit mentioned above is considered within 
the framework of studying the distributions of GRBs in e.g. the \ama\ plane. Indeed, this 
limiting line can account for the absence, in the observed GRB sample with measured $z$ and 
well constrained peak energy (i.e. the bursts used to construct the \ama\ correlation), of bursts with 
intermediate/low peak energy \ep\ and very large \eiso.

The assumed distributions of \G\ and \th\ determine the distribution of 
simulated bursts in the \ama\ plane of Fig. \ref{fg0}. 
We considered two types of distributions for \th\ and \G: 
(A) a power law distribution, i.e. \th\ and \G\  do not assume any preferential value or 
(B) both \th\ and \G\ have peaked distributions (either  broken power law or a log--normal distributions). 

In order to test these two hypothesis we compared the results of our simulations 
with three GRB samples: the complete \sw\ sample of GRBs detected by BAT with measured 
redshifts (S12), the sample of bursts detected by the GBM in the last 2 years 
(Goldstein et al. 2012) and the 4th BATSE catalog of GRBs (Meegan et al. 1997). 
The simulations should reproduce several proprieties of these samples. 
 
While most of  the bright bursts of the \sw\ complete sample of S12 have measured $z$ and provide an 
observational constrain in the rest frame \ama\ and observer frame \epof\ plane (left and right panels of 
Fig.~\ref{fg1},\ref{fg2},\ref{fg3}, respectively), the number count distribution and rate of the \ba\ and GBM 
populations of bursts (mostly without measured $z$) are used as additional constraints since they map the 
faint end of the number count distribution of GRBs.

Our main result is that we cannot reproduce all our observational constraints 
if the \G\ and \th\ distributions are power laws. 
In this case the rate of GBM and BATSE bursts predicted by our simulations 
is a factor $\sim$2 larger than the real one 
and the distribution of the \sw\ simulated bursts in the \ama\ and \epof\ plane 
is inconsistent with the real complete sample of \sw\ bursts. 

Instead, if \th\ and \G\ have broken power law distributions (with peak values 
\th$\sim 4.5^\circ$ and \G$\sim 70$) 
or log--normal distributions (with peak values \th$\sim 4.5^\circ$ and \G$\sim 85$) a 
better agreement between the simulations and the observational constraints is found. 
However, the broken power law 
or log--normal case produce a linear \ama\ correlation due to the assumption that 
the simulated bursts have a \th\ distribution with a unique peak value (see \S4). 
This motivated us to consider the possibility that there is a relation between 
the peak values of the distributions of \th\ and \G. 
G12 found that among GRBs with a \G\ estimate, three new correlations 
are found: \eiso$\propto$\G$^2$,  \liso$\propto$\G$^2$ and \ep$\propto$\G. 
The combination of these correlations with the assumptions that $\theta_{\rm iso}^2$\G=const allows 
to derive the three main empirical correlations of GRBs: the \ama\ correlation, the \yone\ 
correlation and the \ghi\ correlation. 

We therefore assumed that both \th\ and \G\ have log--normal distributions and that a 
relation of the type $\theta_{\rm jet}^m$\G=const  exists between the peak values of their respective 
log--normal distributions. 
We found good consistency between our simulations and 
the observational constraints (Fig. \ref{fg3}) in the case of a log--normal distribution of 
\G\ with central value 90 and logarithmic dispersion of 0.65. 
The distribution of \th\ (also a log--normal) is in this case determined by the relation 
$\theta^{m}_{\rm jet}$\G=const which we find should have $m=2.5$. 
This value is what one obtains by combining the above scaling relations (between \G\ 
and \eiso, \ep) with the \ama\ correlation of the \sw\ complete sample which is \ep$\propto E_{\rm iso}^{0.6}$. 
The existence of a relation  $\theta^{m}_{\rm jet}$\G=const (with $m\sim1$ and const=10--40) is also predicted from recent 
models of magnetically accelerated jets in GRBs (e.g. Tchekhovskoy et al. 2011). 
 
We found that the \G\ distribution that best reproduces all our observational constraints extends to low \G\ values. 
If we cut the \G\ distribution so to exclude such low values of \G\ we cannot reproduce the observed flux and fluence distributions and detection rates of \ba\ and GBM. Therefore, we find that low--\G\ bursts should exists in populations of GRBs
detected by most sensitive detectors. Although a detailed study of the prompt and afterglow properties of these events is out of the scopes of this paper, we can draw some remarks. Apart from their relatively low \ep\ and \liso\ (which are correlated with \G\ as found by G11), the low--\G\ bursts should have a late time afterglow onset (i.e. a few hours for typical parameters, see \S 4.6). Therefore, their prompt emission should not be contaminated by the afterglow while their late time afterglow onset could be one of their distinctive features.  

An immediate consequence of our results is that the large scatter of the \ama\ 
correlation can be interpreted as due to the jet opening angle distribution of GRBs. 
The found inverse relation between \th\ and \G\ implies that bursts with 
the largest bulk Lorentz factors should have a smaller average \th.  
On the other hand, bursts with relatively low average \G\ factors should 
also have, on average, large \th.

Our results depend on the assumption that all bursts have the same \epcom=1.5 keV and 
\egcom=1.5$\times 10^{48}$ erg. Although there could be a dispersion of these values, our 
results still hold if the width of this dispersion is not larger than the dispersion of the observed quantities. 
We note that larger values of \epcom\ and \egcom\ would move the \ep$\propto$\eiso$^{1/3}$ (Eq.~\ref{gammalimit}) 
towards the upper part of the plane of Fig.~\ref{fg0}. As a consequence some of the real GRBs 
of the \sw\ complete sample, would be cut out of the plane because they would lie in the forbidden region (III) of this plane. 
On the other side we could assume lower values of  \epcom\ and \egcom. Since we do not know their real dispersion, we 
tried to assume \epcom=0.15 keV and \egcom=1.5$\times 10^{47}$ (i.e. a factor 10 lower than the values assumed in the 
simulation). Under this different assumption, for the case of log--normal distributions of both \th\ and \G\ and of an 
intrinsic relation between these two parameters, we find that the \th\ distribution is consistent with that found with the fiducial 
values of \epcom\ and \egcom, but with a different distribution of \G. Indeed, in this case we find a mean value 
\G$\sim$ a factor 3 larger than that of the present simulation. Although it is not possible at 
the present stage to constrain the distribution of \epcom\ and \egcom, these results suggest that their dispersion should be 
lower than a factor of $\sim$10.  

Our best simulations allow us to derive the properties of three populations of GRBs: 
those that are pointing to us and that have a peak flux bright enough to enter in the 
\sw\ bright sample (i.e. with the same peak flux threshold adopted for the \sw\ 
complete sample of S12), those that are pointing to us and, finally the full population 
of simulated GRBs, oriented randomly in the Universe (i.e. pointing to us and not). 
The latter is the GRB population that we cannot study on the base of the bursts that we detect.  
The main advantage of our population synthesis code is that we can infer the properties (e.g. 
the \G\ and \th\ distribution and the true GRB rate in this work) of this population of bursts, which is unaccessible 
through the observations.

One immediate consequence of our simulation is the true \ama\ correlation. 
If we consider the PO bursts and if we were in principle able to detect them all, we should find 
a different \ama\ correlation than the one presently reported in the literature. Indeed, the fit of the 
PO bursts in the \ama\ plane of Fig.~\ref{fg3} yields a correlation with slope 0.5 and normalization 
-27.6 while the entire GRB population, the total simulated bursts, have a \ama\ correlation with slope 0.44 and normalization
-20.7.  This is due to the fact that PO bursts tend to populate the lower region of the \ama\ plane (Fig.\ref{fg3}) where the $E_{\rm iso}^{1/3}$ limit cuts their distribution in the \ama\ plane. Therefore,  if we could measure \eiso\ and \ep\ for all the bursts that point to us, we should determine a flatter \ama\ correlation than that observed so far in the high part of the plane with bright bursts.

Our simulation predicts that the bright bursts detected by \sw\ 
should have a mean opening angle of \th$\sim$4.7$^\circ$. 
This value is only a factor 2 smaller than the mean of the entire GRB population that 
we have simulated (which has \th$\sim 8.7^{\circ}$). 
However, from Fig. \ref{fg4} (open blue squares) one can see that if we were 
able to detect fainter GRBs and to measure their jet opening angle, we would obtain 
a mean of 40$^\circ$.  
Intriguingly we note that the present distribution of \th\ measured from the optical 
afterglow break times in a few bursts 
is representative of the \th\ distribution of the entire population of bursts.
This is because the bursts that we have detected so far populate the high region of the \ama\ plane  
where the \th\ distribution can be almost unbiasedly sampled. 
In fact, only the bursts at lower values of \ep\ and \eiso\ are affected by the ``natural bias" 
of 1/\G$>$sin\th.
The low \ep -- low \eiso\ region is where PO bursts concentrate (they 
have large \th\ or small \G, enhancing the probability to point at us).

Our simulation predicts that there are bursts with no jet break, the ones with 1/\G$>$sin\th. 
Their afterglows will never have a jet break since the condition 1/\G$\sim$sin\th\ is never met
 but their afterglow light curve should have a characteristic post--jet break intermediate/steep decay 
slope.
These should be $\sim$6\% of the bursts pointing to 
us and $\sim$2\% of the bursts detected by \sw\ with P$>$\flim. 

According to our best simulation, the mean \G\ of all bursts is $\langle\Gamma_0\rangle=274$.
The \G\ distribution is highly asymmetric and 
there is a considerable difference between its mode (i.e. the peak) the mean and the median. 
The simulated bursts pointing to us corresponding to the \sw\ complete sample have 
$\langle\Gamma_0\rangle=355$. 
These two values are broadly consistent, as explained above, since these bursts populate the upper 
part of the \ama\ plane where the distribution of GRBs is almost free from the ``natural bias". 
Remarkably, if we were able to measure \G\ for all the bursts pointing to us, we would find 
a very low value of the mean of $\langle\Gamma_0\rangle=50$. 
Finally, we have found that the distribution of \G\ that we predict for the \sw\ 
bright sample is consistent with the distribution of \G\ of the GRBs studied in G12.

We can derive from our simulations the true rate of GRBs.  
Previous studies of the GRB rate assumed a unique value of \th, typically 0.2 rad or the {\it observed} distribution of 
\th (e.g. Guetta et al. 2005; Grieco et al. 2012). Our simulations (\S4.5) show that the peak of the intrinsic/global distribution 
of \th\ is  a factor 2 larger than the real intrinsic distribution and has a much wider dispersion (Tab.1).  
Differently from existing GRB rate estimates based on the correction of the isotropic GRB rate 
for an {\it average} beaming factor (e.g. Guetta et al. 2005; Grieco et al. 2012) in 
our simulation the total number of simulated bursts is adjusted in order to reproduce 
the rate of detections of GBM and BATSE. 
Therefore, we have the rate of GRBs as a function 
of redshift independently from the value of \th\ of each single burst. 
If we compare this rate with that of SNIb/c (from Gireco et al. 2012) we find that the 
local rate of GRBs is $\sim$0.3\%. 
Moreover, if we consider the  7\% fraction of  SNIb/c 
which produce Hypernovae events (Guetta \& Della Valle 2007) we find 
that the rate about 4.3\% of local Hypernovae should produce a GRB.

\section*{Acknowledgments}
We thank the referee for comments and suggestions that improved the manuscript. 
We acknowledge ASI I/004/11/0 
and the 2011 PRIN-INAF grant for financial support.


\begin{thebibliography}{}
\bibitem{} Amati, L., Frontera, F., Tavani, M. et al. 2002, A\&A, 390, 81
\bibitem{} Amati, L., Frontera, F., \& Guidorzi, C. 2009, A\&A, 508, 173
\bibitem{} Band, D., Matteson, J., Ford, L. et al. 1993, ApJ, 413, 281
\bibitem{} Band, D. L., \& Preece, R. 2005, ApJ, 627, 319 
\bibitem{} Barbiellini, G., Long0, F., Omodei, N., et al. 2006, NCimB, 121, 1363
\bibitem{} Bosnjak, Z., Celotti, A., Longo, F., Barbiellini, G., 2008, MNRAS, 384, 599
\bibitem{} Burrows D. N., Romano P., Falcone A., et al. 2005, Sci, 309, 1833
\bibitem{} Butler, N. R., Kocevski, D., Bloom, J. S., \& Curtis, J. L. 2007, ApJ, 671, 656 
\bibitem{} Butler, N. R., Kocevski, D., \& Bloom, J. S. 2009, ApJ, 694, 76 
\bibitem{} Collazzi A. C., Schaefer B. E., Goldstein A., Preece R. D., 2012, ApJ, 747, 39
\bibitem{} Dado S., Dar A., De Rujula A., 2007, ApJ, 663, 400
\bibitem{} Eichler, D., \& Levinson, A. 2005, ApJ, 635, 1182
\bibitem{} Falcone A. D., Morris D., Racusin J., et al., 2007, ApJ, 671, 1921
\bibitem{} Firmani C., Ghisellini G., Ghirlanda G., Avila-Reese V., 2005, MNRAS, 360, L1
\bibitem{} Firmani C., Cabrera J.I., Avila-Reese V., Ghisellini G., Ghirlanda G., et al., 
          2009,  MNRAS, 393, 1209
\bibitem{} Frail D. A., Kulkarni S. R., Sari R., et al, 2001, ApJ,  526, L55
\bibitem{} Ghirlanda, G., Ghisellini, G., Lazzati, D. 2004, ApJ, 616, 331
\bibitem{} Ghirlanda, G., Ghisellini, G., Firmani, C. 2005, MNRAS, 361, L10
\bibitem{} Ghirlanda, G., Nava, L., Ghisellini, G., Firmani, C., Cabrera, J. I., 2008, MNRAS, 387, 319
\bibitem{} Ghirlanda G., Nava L., Ghisellini G., Firmani C., 2007, A\&A, 466, 127
\bibitem{} Ghirlanda, G.; Nava, L.; Ghisellini, G.; Celotti, A.; Firmani, C.,  2009, A\&A, 496, 585
\bibitem{} Ghirlanda, G., Nava, L.; Ghisellini G., 2010, A\&A, 511, 43
\bibitem{} Ghirlanda, G., Ghisellini G., Nava, L., 2011, MNRAS, 418, L109
\bibitem{} Ghirlanda, G., Ghisellini G., Nava, L., 2011a, MNRAS, 410, L97
\bibitem{} Ghirlanda, G., Nava, L.; Ghisellini G., et al., 2012, MNRAS, 420, 483 (G12)
\bibitem{} Ghirlanda, G. Ghisellini G., Nava, L. et al., 2012b, MNRAS, 422, 2553
\bibitem{} Ghisellini G., Nardini M., Ghirlanda G., Celotti A., 2009, MNRAS, 393, 253
\bibitem{} Ghisellini G., Ghirlanda G., Nava L., Celotti A., 2010, MNRAS, 403, 926
\bibitem{} Giannios, D., \& Spruit, H. C. 2007, A\&A, 469, 1 
\bibitem{} Giannios, D., 2012, MNRAS, 422, 3092
\bibitem{} Goldstein A., Burgess J. M., Preece R. D., et al., 2012, ApJS, 199, 19 
\bibitem{} Grieco V., et al., 2012, MNRAS in press, arXiv:1204.2417
\bibitem{} Grupe D., Gronwall C.; Wang X.-Y., et al., 2007, ApJ, 662, 443 
\bibitem{} Guetta D., Piran T., Waxman E., 2005, ApJ, 619, 412

\bibitem{} Guetta D. \&  Della Valle M., 2007, A\&A, 461, 95
\bibitem{} Hopkins A. M., Beacom J. F., 2006, ApJ, 651, 142 
\bibitem{} Kaneko, Y., Preece, R. D., Briggs, M. S. et al. 2006, ApJS, 166, 298
\bibitem{} Kocevski D. \& Butler N., 2008, ApJ, 680, 531
\bibitem{} Kocevski, D., 2012, ApJ, 747, 146
\bibitem{} Komissarov S. S., Vlahakis N., Koenigl A., 2010, MNRAS, 407, 17
\bibitem{} Krimm, H. A., Yamaoka, K., Sugita, S., et al. 2009, ApJ, 704, 1405
\bibitem{} Lamb, D. Q., Donaghy, T. Q., \& Graziani, C. 2005, ApJ, 620, 355 
\bibitem{} Lazzati D., Morsony B. J. \& Begelman M. C., 2011, ApJ, 732, 34
\bibitem{} Lenvinson, A., \& Eichler, D. 2005, ApJ, 629, L13 
\bibitem{} Li, Li-Xin, 2008, MNRAS, 388, 1487
\bibitem{} Meegan C. A., Paciesas W. S., Pendleton G. N., et al., 1998, AIPC, 428, 3 
\bibitem{} Mundell C. G., Melandri A., Guidorzi C., et al., 2007, ApJ, 660, 489
\bibitem{} Nakar, E. \& Piran, T. 2005, MNRAS, 360, L73
\bibitem{} Nava L., Ghisellini G., Ghirlanda G., et al., 2006, 450, 471
\bibitem{} Nava, L., Ghirlanda, G., Ghisellini, G, Firmani, C. 2008, MNRAS, 391, 639
\bibitem{} Nava, L., Ghirlanda, G., Ghisellini, G, et al., 2011a, A\&A, 530, 21
\bibitem{} Nava, L.,  Ghirlanda, G.; Ghisellini, G.; Celotti, A., 2011b,MNRAS, 415, 3153 
\bibitem{} Nava, L., Salvaterra, R., Ghirlanda, G., et al., 2012 (N12), MNRAS, 421, 1256
\bibitem{} Paciesas W. S., Meegan C. A., von Kienlin A., et al., 2012, ApJS, 199, 18
\bibitem{} Panaitescu A., \& Kumar P., 2001, ApJ, 560, L49
\bibitem{} Panaitescu A., 2009, MNRAS, 393, 1010
\bibitem{} Piran, T., 1999, PhR, 314, 575
\bibitem{} Racusin J. L., Liang E. W., Burrows D. N., et al., 2009,  ApJ, 698, 43
\bibitem{} Rees, M., \& Meszaros, P. 2005, ApJ, 628, 847
\bibitem{} Ryde, F., Bjornsson, C., Kaneko, Y., et al. 2006, ApJ, 652, 1400
\bibitem{} Sakamoto T., Barthelmy S. D., Baumgartner W. H., et al., 2011, ApJS, 195, 2
\bibitem{} Salvaterra R., Campana S., Vergani S. D., et al., 2012, ApJ, 749, 68
\bibitem{} Shahmoradi A., \& Nemiroff R. J., 2011, MNRAS, 411, 1843 
\bibitem{} Soderberg A. M., Kulkarni S. R., Nakar E., et al., 2006, Nature, 422, 1014
\bibitem{} Toma, K., Yamazaki, R., \& Nakamura, T. 2005, ApJ, 635, 481 
\bibitem{} Tchekhovskoy A., McKinney J. C., Narayan R., 2009, ApJ, 699, 1789
\bibitem{} Thompson, C. 2006, ApJ, 651, 333
\bibitem{} Thompson, C., Meszaros, P., \& Rees M. J. 2007, ApJ, 666, 1012
\bibitem{} Van Eerten, H., Zhang, W., MacFadyen A., 2010, ApJ, 772, 235
\bibitem{} Van Eerten, H., Meliani, Z, Wijers, R. A. M., Kippens, R., 2011, MNRAS, 410, 2016
\bibitem{} Yamazaki, R., Ioka, K., \& Nakamura, T. 2004, ApJ, 606, L33
\bibitem{} Yonetoku, D., Murakami, T., Nakamura, T. et al. 2004, ApJ, 609, 935

\end{thebibliography}
\end{document}